\documentclass[12pt,aps,prd,nofootinbib,superscriptaddress]{revtex4-2}
\usepackage[utf8]{inputenc}
\usepackage{amsmath,amssymb}
\usepackage[T1]{fontenc}
\usepackage{mathtools}
\usepackage{fontawesome}
\usepackage[colorlinks,pdfusetitle]{hyperref}
\hypersetup{allcolors=[rgb]{1,0.56,0}}

\usepackage{mathrsfs}
\usepackage{appendix}
\usepackage{float}
\usepackage{xspace}

\newcommand{\figref}{Fig.~\ref}
\newcommand{\forref}[1]{Eqn.~(\ref{#1})}
\newcommand{\zbl}[1]{\textcolor{black}{#1}}
\newcommand{\zblb}[1]{\textcolor{black}{#1}}
\newcommand{\zblc}[1]{\textcolor{black}{#1}}
\newcommand{\zbld}[1]{\textcolor{black}{#1}}

\newcommand{\cevns}{CE$\nu$NS\xspace}
\newcommand{\tonyear}{$\text{ton}\times\text{years}$\xspace}
\newcommand{\Hzero}{\text{H}_0}
\newcommand{\Hone}{\text{H}_1}

\begin{document}
\title{Asymptotic Analysis on Binned Likelihood and Neutrino Floor}

\author{Jian Tang}%
 \email{tangjian5@mail.sysu.edu.cn}
\author{Bing-Long Zhang}
 \email{zhangblong@mail2.sysu.edu.cn, to whom all correspondence should be addressed.}
\affiliation{
School of Physics, Sun Yat-Sen University, Guangzhou 510275, China
}%
\date{\today}

\begin{abstract}
Observations of suspected coherent elastic neutrino-nucleus scatterings by dark matter direct detection experiments highlight the need for an investigation into the so-called "neutrino floor". We focus on the discovery limit, a statistical concept to identify the neutrino floor, and analyze the asymptotic behaviour of the profile binned likelihood ratio test statistic where the likelihood is constructed by variate from events in each bin and pull terms from neutrino fluxes. To achieve the asymptotic result, we propose two novel methods: i)  Asymptotic-Analytic method, which furnishes the analytic result for large statistics, is applicable for more extra nuisance parameters, and enables the identification of the most relevant parameters in the statistical analysis; ii) Quasi-Asimov dataset, which is analogous to Asimov dataset but with improved speed. Applying our methods to the neutrino floor, we significantly accelerate the computation procedure compared to the previous literature, and successfully address cases where Asimov dataset fails. Our derivation on the asymptotic behavior of the test statistic not only facilitates research into the impact of neutrinos on the search for dark matter, but may also prove relevant in similar application scenarios.
~\href{https://github.com/zhangblong/AsymptoticAnalysisAndNeutrinoFog}{\large\faGithub}
% purpose, method, result, conclusion
\end{abstract}

\maketitle

\section{Introduction}
\label{introduction}

% Explain the importance of neutrino and DM. The CEvNS was confirmed. Necessary to quantificationally decribe the competition relationship between neutrinos and DM.
The nature of dark matter (DM) is one of most alluring fields in physics but still remains unknown with several decades of efforts on searching for DM~\cite{Schumann:2019eaa,Gaspert:2021gyj,Cirelli:2010xx}. 
From the perspective of DM direct detection (DD) experiments, however, constraints on DM, particularly a popular candidate -- weakly interacting massive particles (WIMPs), have been improved through the dedicated efforts of DD experiments, which inspires us to pursue increasingly sensitive techniques for detecting recoil signals generated by WIMPs. 
The recent discovery~\cite{COHERENT:2017ipa} of the coherent elastic neutrino-nucleus scattering (\cevns) process by COHERENT collaboration represents a recent milestone in this field. DM DD experiments, such as XENON and PandaX~\cite{XENON:2020gfr,PandaX:2022aac}, have announced their observations of suspected solar neutrino signals generated by \cevns.
However, as sensitivities increase, the presence of irreducible neutrino backgrounds poses a challenge to the search for WIMPs, given the similarities between their respective signals. As such, it is essential to quantitatively evaluate the impact of neutrinos on direct searches for WIMPs.

% State the history of neutrino floor and recent progress. 
Neutrino floor, proposed to systematically quantify the discovery potential of searching for WIMPs in the presence of neutrino backgrounds~\cite{Billard:2013qya}, 
is presented by a curve on the parameter space of WIMP-nucleon cross section versus WIMP mass, below which WIMPs cannot be detected significantly. Formally, the neutrino floor can be defined by discovery limits~\cite{Billard:2011zj}: the minimum cross section required for an experiment to have a 90\% probability or 90\% confidence level (C.L.) to detect a WIMP signal with a 3$\sigma$ significance~\cite{Ruppin:2014bra}. Obtaining the neutrino floor traditionally requires Monte Carlo (MC) pseudo-experiments which can be time-consuming. 
Additional studies on the neutrino floor have explored various factors such as astrophysical uncertainties~\cite{OHare:2016pjy}, altering the WIMP-nucleon interaction with the non-relativistic effective field theory formalism~\cite{Dent:2016iht} and modifying neutrino backgrounds with neutrino related new physics~\cite{Gonzalez-Garcia:2018dep,Bertuzzo:2017tuf,AristizabalSierra:2021kht}, etc. There are several studies on how to overcome the neutrino floor, including combining data from experiments using different targets~\cite{Ruppin:2014bra}, taking advantage of the annual modulation~\cite{Davis:2014ama} or the diurnal modulation~\cite{Sassi:2021umf},  and adopting more realistic directional detection strategy~\cite{OHare:2015utx,OHare:2017rag,OHare:2020lva,Grothaus:2014hja,Mayet:2016zxu,Franarin:2016ppr,Vahsen:2020pzb,Vahsen:2021gnb} with lower statistics.
Recently, a new definition on the neutrino floor has been proposed, which does not rely on choices on detector thresholds and exposures~\cite{OHare:2021utq}. Instead, it uses the gradient of the discovery limit cross section with respect to the exposure to measure the marginal utility on the WIMP searches, resulting in the definition of the "neutrino fog". The neutrino floor is then identified as a boundary of the neutrino fog, representing the transition from statistical to systematical limits~\cite{Akerib:2022ort}. To reduce computational costs, the confidence level in the previous definition was adjusted to 50\% and Asimov dataset~\cite{Cowan:2010js} was used to obtain the median discovery limit.

% Raise problems about the statistic method and computation cost. State our treatment about these problems. 
There appears to be some unresolved issues for determining the neutrino floor: i) the statistic methodology for the discovery limit lacks further investigation; ii) \zblb{the calculation of the neutrino floor necessitates substantial computational resources, and enhancing the stability of the algorithm is imperative}; iii) Asimov dataset can not resolve cases that some specific degrees of freedom (DOFs) such as astrophysical uncertainties are considered, and iv) MC pseudo-experiments increasingly cost time when more parameters are considered. 
Thus, in this paper, we infer the asymptotic behaviour of the profile binned likelihood ratio test statistic (PBLRTS) where the likelihood is constructed by different variables. The similar case, where the likelihood is constructed by variables following the same distribution, has been worked out~\cite{Wilks:1938dza,Wald1943TestsOS}.
Subsequently, we propose two new methods to achieve the asymptotic result: i) Asymptotic-Analytic method which provides analytic results for large statistics, is general for more extra nuisance parameters and affords a way to determine the most relevant parameters in the statistical analysis; and ii) Quasi-Asimov dataset analogous with but faster than Asimov dataset method. 
\zblb{By applying our methods to the neutrino floor, we significantly accelerate the computation procedure used in previous literature, investigate the neutrino fog taking into account the uncertainty of the weak mixing angle, and handle the case when a specific DOF from the astrophysical uncertainty is considered, which can not be addressed by Asimov dataset.} Moreover, the most relevant parameters from neutrino fluxes can be obtained by Asymptotic-Analytic method, so that MC pseudo-experiments can be boosted by neglecting non-dominant contributions from neutrino sources. Furthermore, our proposed methods are not only applicable for the neutrino floor and fog~\cite{OHare:2020lva,AristizabalSierra:2021kht,OHare:2021utq}, but also might be feasible in searching for the bump of the diffuse flux of high-energy cosmic neutrinos in IceCube~\cite{Fiorillo:2022rft}, studying the impact of DM on DSNB in Hyper-Kamiokande experiment~\cite{Bell:2022ycf}, detecting time-varying DM signals with Paleo-Detectors~\cite{Baum:2021chx} and analyzing the daily modulation on dark photon~\cite{Caputo:2021eaa}.

This paper is organized as follows. In section \ref{statistic}, we provide some key process in obtaining the asymptotic behaviour of PBLRTS. In section \ref{neutrino}, we describe the statistic method behind the neutrino floor and provide the result from our new methods. In addition, Asymptotic-Analytic method is used to analytically explain the evolution of the discovery limit cross section with the exposure. \zblb{In section \ref{ANP}, we provide the neutrino fog and floor with considering the uncertainty of the weak mixing angle,and afford the modified discovery limit curve for experimental configurations with considering nuisance parameter from the velocity of the local standard of rest.} Finally, in section \ref{conclusion} we draw our conclusions and outlooks.

\section{Asymptotic Behaviour of Profile Binned Likelihood Ratio Test Statistic}
\label{statistic}
\subsection{Derivation on Asymptotic Distribution}
\label{Derivation}
Our notation adheres to the conventions described in the book on large sample theory~\cite{Ferguson2017}.
To begin with, we recall a theorem concerning the large sample distribution of the likelihood ratio test statistic (LRTS)~\cite{Wilks:1938dza,Wald1943TestsOS}. In the context of the likelihood ratio test, we consider a parameter set with k DOFs denoted by $\boldsymbol{\theta} \in \Theta \subset \mathbb{R}^k$, where r ($1\leq r \leq k$ ) is the number of parameters of interest and (k-r) is the number of nuisance parameters. The likelihood function $L(\boldsymbol{\theta})$ is constructed by variables following the same distribution. The likelihood ratio test provides a general way for discriminating the null hypothesis $\Hzero:~\boldsymbol{\theta}^0 \in \Theta_0$ versus the alternative hypothesis $ \Hone:~ \boldsymbol{\theta}^1 \in \Theta - \Theta_0$, where $\Theta_0 \subset \Theta$. Defining the log likelihood function $l(\boldsymbol{\theta})\equiv \log L(\boldsymbol{\theta})$ and $\lambda \equiv \frac{L(\boldsymbol{\theta}^*)}{L(\hat{\boldsymbol{\theta}})}$, the null hypothesis is rejected if the likelihood ratio test statistic $-2\ln \lambda = -2[\ln l(\boldsymbol{\theta}^*)-\ln l(\hat{\boldsymbol{\theta}})]$ is larger than our expectation, where $\boldsymbol{\theta}^*$ and $\hat{\boldsymbol{\theta}}$ represent the maximum-likelihood estimator (MLE) over $\Theta_0$ and $\Theta$, respectively. In general, $\Theta_0$ is a (k-r)-dimensional subspace of $\Theta$, with r constraint conditions given by: $g_i(\boldsymbol{\theta})=0$ ($1\leq i\leq r$). Under the assumption of large samples, if the true model $\boldsymbol{\theta}^\prime$ satisfies $\Hzero: \boldsymbol{\theta}^0$, the quantity $-2\ln \lambda$ follows the chi-square distribution with r DOFs, denoted by $-2\ln \lambda \sim \chi_r^2$. If  $\boldsymbol{\theta}^\prime$ deviates from $\boldsymbol{\theta}^0$, 
 then $-2\ln \lambda \sim \chi^2_{r}(\phi)$, where $\phi$ is the non-central parameter related to $\boldsymbol{\delta}=\boldsymbol{\theta}^\prime-\boldsymbol{\theta}^0$. 

However, the aforementioned theorem should be modified to be applied to our scenario, where the binned likelihood comprises variables that follow different distributions and are accompanied by pull terms. Consequently, it becomes imperative to extend the theorem scope. To this end, we invoke the Lyapunov central limit theorem (CLT)~\cite{Krishna2006}  to approximate some quantities to its expectation or acquire the asymptotic distribution of some quantities , which are constructed by random variate that might not be from the same distribution, in our proof. Moreover, our derivation in the following is based on the assumption that the likelihood function is smooth and its derivatives are bounded.

\zbl{Here we simply expound upon the test statistic's asymptotic formula, and please refer to Appendix~\ref{appendix1} for the detailed derivation. The asymptotic formula is constructed by the derivative of the logarithmic likelihood function. One is the first derivative at  $\boldsymbol{\theta}^\prime$, i.e.,  $\dot{l}(\boldsymbol{\theta}^\prime)$, which follows a multivariate normal distribution with a mean vector $\boldsymbol{\mu}$ and a covariance matrix $\mathbf{V}$, denoted by $\mathscr{N}(\boldsymbol{\mu}, \mathbf{V})$. The other is the expectation of the second derivative at  $\boldsymbol{\theta}^\prime$, i.e.,  $E(\ddot{l}(\boldsymbol{\theta}^\prime))$. For convenience, we define $\mathscr{F} \equiv -E(\ddot{l}(\boldsymbol{\theta}^\prime))$ and a matrix $\mathbf{H}$ which is also constructed by $E(\ddot{l}(\boldsymbol{\theta}^\prime))$, and details are provided in Appendix \ref{appendix1}. Utilizing the Taylor's expansion, the Lyapunov CLT and some properties of quantities, the asymptotic formula for the test statistic can be written as:
\begin{equation} \label{eqn:StatisticDistribution}
-2\ln \lambda \approx Z^T \mathbf{V}^{\frac{1}{2}}[\mathscr{F}^{-1}-\mathbf{H}]
\mathbf{V}^{\frac{1}{2}} Z \,,
\end{equation}
where $Z \sim \mathscr{N}(\mathbf{V}^{-\frac{1}{2}} (\boldsymbol{\mu}+\mathscr{F} \boldsymbol{\delta}), I)$ is a vector. To make the result more explicit, we perform a diagonalization: $-2\ln \lambda \approx Z^T U^T \Lambda U Z=Y^T \Lambda Y$, where $\Lambda$ is a diagonal matrix, $U$ is a special orthogonal matrix and $Y = U Z \sim \mathscr{N}(U \mathbf{V}^{-\frac{1}{2}}(\boldsymbol{\mu}+\mathscr{F} \boldsymbol{\delta}), I)$. As a consequence, $-2\ln \lambda$ asymptotically follows the distribution of a sum over several non-central chi-square variate with varying weights: 
\begin{equation}
    -2\ln \lambda \approx \sum_i a_i Y_i^2\,,
\end{equation}
where $a_i$ is the $i^{th}$ component of the diagonal term in $\Lambda$, and $Y_i^2 \sim \chi^2_1(\phi_i)$ where $\phi_i$ is the non-central parameter. For instance, in the case presented in the next section where $\Hzero$ is true, we have $\boldsymbol{\delta}=0$ and $Y\sim \mathscr{N}(0, I)$. Consequently, $-2\ln \lambda \sim \chi^2_1$, aligning with Wilk's theorem.} \footnote{Generally speaking, explicit forms for $a_i$ and $\phi_i$ are not readily available, and should be computed numerically by evaluating  $\Lambda$ and $Y$.} 

Moreover, it is noteworthy to observe that the entries within the matrix $U$ signify the impact exerted by each parameter. As a result, one can minimize the cost of MC pseudo-experiments by selectively incorporating only the most relevant parameters, and the criteria can be derived from the matrix $U$ in our Asymptotic-Analytic method. This technique has been applied in our MC realizations, which is provided in Appendix \ref{MCCheck}.

\subsection{Asymptotic-Analytic Method}
\label{AA}
The Asymptotic-Analytic Method has been presented previously, but it did not include a discussion on how to obtain the values of  $\boldsymbol{\mu}$, $\mathbf{V}$ and $ \mathscr{F}$. Here, we provide a concrete application.

For simplicity, we only consider the case of signal discovery where there is only one parameter of interest and all parameters are normalized to unity. For the discovery of a signal, $\theta_1$ represents the signal strength, while $\theta_i~(i\geq 2)$ are nuisance parameters. \zblc{In many literature, the signal strength is always denoted by $\mu$, while we use $\theta_1$ here for convenience. Under the null hypothesis $\Hzero:~\boldsymbol{\theta}^0$, we have $\theta_1^0=0, ~\theta_i^0=1(i\geq 2)$. Here, the bold symbol represents a vector where the superscript marks the model, and the normal symbol like $\theta_i^0$ represents the i-th component of $\boldsymbol{\theta}^0$.} Under the alternative hypothesis $\Hone:~\boldsymbol{\theta}^1$, we have $\theta_i^0=1(i\geq 1)$. We adopt the binned likelihood \zblc{with N bins and M parameters} as shown below:
\begin{equation}\label{eqn:Likelihood}
L(\theta_1, \theta_2,\dots,\theta_m ) =
\prod_{i=1}^N \mathscr{P}(n_i|\theta_1 s_i + b_i)   \;\;
\prod_{j=2}^M \mathscr{N}_{\theta_j}(1, \sigma_j)  \;,
\end{equation}
where $n_i,~s_i$ and $b_i$ are observed, signal and background events in $i^{th}$ bin, respectively. The variable $n_i$ follows a Poisson distribution with a mean of the expected value $v_i=\theta_1 s_i + b_i$, i.e., $\mathscr{P}(n_i|v_i)=v_i^{n_{j}} e^{- v_i }/n_{j}!$. The quantities $s_i$ and $b_i$ are dependent on the nuisance parameters  $\theta_i~(i\geq 2)$, which follows a normal distribution with a mean of 1 and a standard deviation $\sigma_j$: $\mathscr{N}_{\theta_j}(1, \sigma_j)=\frac{1}{\sqrt{2\pi} \sigma_j} \exp{(-\frac{(\theta_j-1)^2}{2 \sigma_j^2})}$.

Furthermore, we obtain the expressions for the expectation vector $\boldsymbol{\mu}$, the variance matrix $\mathbf{V}$ of the first derivative of $l(\boldsymbol{\theta})$ at $\boldsymbol{\theta}^\prime$: $\dot{l}(\boldsymbol{\theta}^\prime)$ and the expectation of the second derivative with a minus sign $\mathscr{F}$ \zblc{as follows:
\begin{equation}\label{eqn:EandV}
\begin{aligned}
&\mu_\alpha = \left\{\begin{array}{cc}
    \sum_{i j} \frac{1}{2}\frac{\partial^2 v_i}{\partial \theta_j^2} \frac{\sigma_j^2}{v_i} \frac{\partial v_i}{\partial \theta_\alpha}\,, & \alpha=1\,, \\
    \sum_{i j} \frac{1}{2}\frac{\partial^2 v_i}{\partial \theta_j^2} \frac{\sigma_j^2}{v_i} \frac{\partial v_i}{\partial \theta_\alpha} - \frac{\theta_\alpha - \theta_\alpha^\prime}{\sigma_\alpha^2}\,, & 2\leq \alpha \leq M\,,
\end{array}\right.
\\
&\mathscr{F}_{\alpha \beta} = \left\{\begin{array}{cc}
   -\sum_{i j} \frac{1}{2}\frac{\partial^2 v_i}{\partial \theta_j^2} \frac{\sigma_j^2}{v_i} \frac{\partial^2 v_i}{\partial \theta_\alpha \partial \theta_\beta} 
+\sum_{i j} \frac{1}{v_i}(1+\frac{1}{2}\frac{\partial^2 v_i}{\partial \theta_j^2} \frac{\sigma_j^2}{v_i}) \frac{\partial v_i}{\partial \theta_\alpha} \frac{\partial v_i}{\partial \theta_\beta} \,, & \alpha=1~\text{or}~\beta=1\,, \\
    -\sum_{i j} \frac{1}{2}\frac{\partial^2 v_i}{\partial \theta_j^2} \frac{\sigma_j^2}{v_i} \frac{\partial^2 v_i}{\partial \theta_\alpha \partial \theta_\beta} 
+\sum_{i j} \frac{1}{v_i}(1+\frac{1}{2}\frac{\partial^2 v_i}{\partial \theta_j^2} \frac{\sigma_j^2}{v_i}) \frac{\partial v_i}{\partial \theta_\alpha} \frac{\partial v_i}{\partial \theta_\beta} 
+\delta^\alpha_\beta \frac{1}{\sigma_{\alpha}^2}, & 2\leq \alpha,\beta \leq M\,,
\end{array}\right.
\\
&V_{\alpha \beta} = \sum_{i j k l} \left(\frac{1}{v_i} \delta^i_j+\frac{1}{2}\frac{\partial^2 v_i}{\partial \theta_k^2} \frac{\sigma_k^2}{v_i^2}\delta^i_j+
 \frac{\partial v_i}{\partial \theta_k} \frac{\partial v_j}{\partial \theta_k} \frac{\sigma_k^2}{v_i v_j}-
\frac{1}{4} \frac{\partial^2 v_i}{\partial \theta_k^2}
\frac{\partial^2 v_j}{\partial \theta_l^2} \frac{\sigma_k^2 \sigma_l^2}{v_i v_j}
 \right) \frac{\partial v_i}{\partial \theta_\alpha} \frac{\partial v_j}{\partial \theta_\beta},~1\leq \alpha,\beta \leq M\,,
\end{aligned}
\end{equation}
where $\delta^i_j$ and $\delta^\alpha_\beta$ are the Kronecker delta symbols.}
Kindly note that the aforementioned quantities must be assessed at $\boldsymbol{\theta}$, which can be either $\boldsymbol{\theta}^0$ or $\boldsymbol{\theta}^1$. For more details, please refer to Appendix \ref{appendix2}.

After completion of the aforementioned step, we can proceed to compute the asymptotic distribution of $-2\ln \lambda$. This procedure is what we called Asymptotic-Analytic method.
\zblb{Furthermore, we present a rather simple case here for better understanding. In this case, there is only one nuisance parameter considered to modify the background, and the expected value $v_i=\theta_1 s_i + \theta_2 b_i$. According to \forref{eqn:EandV}, we have:
$$
\boldsymbol{\mu}=0,\quad 
\mathscr{F} = \left(\begin{array}{cc}
    \sum_i \frac{s_i^2}{v_i} & \sum_i \frac{s_i b_i}{v_i} \\
    \sum_i \frac{s_i b_i}{v_i} & \sum_i \frac{b_i^2}{v_i} + \frac{1}{\sigma_2^2}
\end{array}\right),\quad
\mathbf{V} = \left(\begin{array}{cc}
    \sum_i (\frac{1}{v_i}+\frac{s_is_j}{v_iv_j}\sigma_2^2)s_i^2 & \sum_i (\frac{1}{v_i}+\frac{s_is_j}{v_iv_j}\sigma_2^2)s_i b_i \\
    \sum_i (\frac{1}{v_i}+\frac{s_is_j}{v_iv_j}\sigma_2^2)s_i b_i & \sum_i (\frac{1}{v_i}+\frac{s_is_j}{v_iv_j}\sigma_2^2)b_i^2
\end{array}\right)\,.
$$
However, it is not feasible to determine the asymptotic distribution of the statistic analytically. As demonstrated in Section~\ref{Analytic} utilizing a numerical method, we ascertain that the test statistic follows  a chi-square distribution when $\Hzero$ is true. When we ask $\theta_1$ for being positive, the test statistic should follows a distribution of $\frac{1}{2}[ \delta(0)+\chi^2_1]$. If $\Hone$ is true, $-2\ln \lambda \sim \chi^2_1(\phi)$, where $\phi$ can be written as:
\begin{equation} \label{eqn:SpecificPhi}
\phi= \sum_i \frac{s_i^2}{s_i+b_i} - \frac{(\sum_i \frac{s_i b_i }{s_i+b_i})^2}{\sum_i \frac{b_i^2 }{s_i+ b_i}+\frac{1}{\sigma_2^2}} \,,
\end{equation}
according to \forref{eqn:phiSimple}.
}
\subsection{Quasi-Asimov dataset Method}
\label{QA}
The Asimov dataset~\cite{Cowan:2010js} was introduced as a means to readily obtain the median values of  the test statistic like  $-2\ln \lambda$ mentioned previously. In this scenario of hypothesis testing, the observations are precisely aligned with their expected values: $n_i=v_i$, where $v_i$ is determined by $\boldsymbol{\theta}$,  while the nuisance parameters in the normal distribution are fixed at $\boldsymbol{\theta}^\prime$. The Asimov dataset is considered an effective approximation method in cases where the sample size is sufficiently large, and is commonly employed in some literature~\cite{AristizabalSierra:2021kht,OHare:2020lva,OHare:2021utq,Fiorillo:2022rft,Bell:2022ycf,Caputo:2021eaa,Baum:2021chx}.

The most time-consuming aspect of implementing Asimov dataset is the numerical search for the MLE  $\boldsymbol{\theta}^*$ for $\Hzero$. However, \forref{eqn:proof3} provides a simple yet effective approach to approximate :
\begin{equation}
\label{eqn:phi2}
\boldsymbol{\theta}^*  \approx \boldsymbol{\theta}^\prime-[1- H \mathscr{F}] \boldsymbol{\delta}\,.
\end{equation}
Hence, we can render the acquisition of the test statistic's median feasible without incurring significant time expense. Note that $\boldsymbol{\theta}^\prime$ is consistently set to $\boldsymbol{\theta}^1$, as the requirement for Quasi-Asimov dataset method is to attain the statistic's median assuming the alternative hypothesis is genuine. We call this procedure Quasi-Asimov dataset method..

However, an issue remains with both Asimov dataset and Quasi-Asimov dataset method. In the signal discovery scenario, Asimov dataset fails when extra nuisance parameters only involve $s_i$. As $\theta^0_1=0$, the extra parameters are constrained to their true values,  resulting in a result that is unaltered by their presence. For example, suppose a new nuisance parameter, $\theta_{k+1}$, is introduced, with a mean of 1 and standard deviation of $\sigma_{k+1}$, and $\theta^0_{k+1}=\theta^1_{k+1}=1$. Then  $\theta^0_1=0$ and the Gaussian term with $\sigma_{k+1}$ compels $\theta^*_{k+1}=1$. Consequently, the median of the test statistic remains unchanged regardless of the inclusion of $\theta_{k+1}$, which is not our expectation. Fortunately, Asymptotic-Analytic method can handle this issue.

\section{Neutrino Floor and Fog}
\label{neutrino}
\subsection{Recoil Spectrum}
In this section, we shall commence by delving into the response of WIMP and neutrinos in the detector as inputs for the neutrino floor. For the sake of simplicity, only the spin-independent WIMP-nucleon interaction is considered here. The differential event rates of WIMP~\cite{Lewin:1995rx} in the detectorcan be expressed as follows:
\begin{equation}
	\frac{dR_\text{WIMP}}{dE_r}=\frac{\rho_0 A^2}{2 m_\chi \mu_N^2}\sigma_0 F^2(E_r) \int_{v_{min}(E_r)}^{v_{esc}}{\frac{f(v,v_0)}{v}d^3v}\,,
\end{equation}
where  $E_r$ is the recoil energy, $m_\chi$ is the mass of WIMP, $\sigma_0$ is the spin-independent WIMP-nucleon cross section, $\mu_N$ is the WIMP-nucleon reduced mass, $A$ is the atom number of the target nucleus, and $F(E_r)$ is the nuclear form factor generally presented by the Helm form~\cite{Helm:1956zz}. The other parameters are taken from the standard halo model (SHM)~\cite{Drukier:1986tm,Evans:2018bqy}:  the WIMP density surrounding the Earth $\rho_\chi = 0.3~\mathrm{GeV / cm}^3$ , the  circular velocity of the Local Standard of Rest (LSR) $v_0=220~\mathrm{km/s}$, the escape velocity of the Milky Way $v_{esc}=544~\mathrm{km/s}$, and $f(v,v_0)$ is the velocity distribution of WIMP. The  quantity $v_{min}(E_r)$ represents the minimum speedat which a WIMP may cause the recoil energy $E_r$, as limited by the kinematics.

Integrating the differential \cevns cross section multiplied by the neutrino flux, we can derive the recoil spectrum for \cevns:
\begin{equation}
	\frac{d R_\nu}{d E_{r}}=\frac{1}{m_N} \sum_i \int_{E_{\nu}^{\min}(E_r)} \frac{d \Phi_i}{d E_{\nu}}\frac{d \sigma_{\nu N}\left(E_{\nu}, E_{r}\right)}{d E_{r}} d E_{\nu}
	\,,
\end{equation}
where $m_N$ is the nuclear mass, $E_{\nu}^{\min}(E_r)$ is the minimum neutrino energy to generate $E_r$, $\frac{d \Phi_i}{d E_{\nu}}$ is the neutrino flux from the source labelled as $i$, and $\frac{d \sigma_{\nu N}\left(E_{\nu}, E_{r}\right)}{d E_{r}}$ is the differential \cevns cross section that can be well described by the standard model. There are numerous neutrino sources causing recoil events in DM detectors and we adopt the same neutrino flux model as in Table I of Ref.~\cite{OHare:2020lva}. Here, we briefly introduce some of the main neutrino sources.

Solar neutrinos are the principal source of \cevns events, which pose an obstacle to the search for the $\mathscr{O}(10)$ GeV-scale WIMP. These neutrinos stem from nuclear fusion reactions such as the pp Chains and the CNO Cycle with energies less than about 13 MeV. They have been meticulously comprehended with the standard solar mode~\cite{Vitagliano:2019yzm}. As the standard solar model is subject to various observational constraints, the parameters governing the model become increasingly precise. This precision allows for the prediction of uncertainties in solar neutrino fluxes. Experimental measurements of $^8$B neutrinos have achieved a high level of precision, resulting in a minimum uncertainty of 2\% over solar neutrinos.
Atmospheric neutrinos produced by cosmic rays interacting with the Earth atmosphere are more energetic but more rare. The atmospheric neutrino flux can be computed through simulations, and the recommended theoretical uncertainty is set at 20\%.
The diffuse supernova neutrino background (DSNB) originates from the cosmological history of core-collapse supernovae in the visible universe with energies roughly 10$\sim$25 MeV at Earth. Due to our incomplete understanding of the DSNB, we set the uncertainty associated with it at 50\%.

\subsection{Statistic Method and Results}
The statistic method about the discovery of a positive signal~\cite{Cowan:2010js} aligns with our discourse on the neutrino floor. As outlined in the Section \ref{statistic}, the null hypothesis $\Hzero$ represents the neutrino background-only model, while the alternative hypothesis $\Hone$ represents the WIMP+neutrino model. The likelihood ratio is given by:
\begin{equation}
\lambda(\theta_1) = \frac{ L(\theta^0_1,
\hat{\hat{\theta}}_2,\dots,\hat{\hat{\theta}}_M) } {L(\hat{\theta}_1, \hat{\theta}_2,\dots,\hat{\theta}_M) } \,,
\end{equation}
where $\theta^0_1 = 0$ represents the signal strength in the background-only model, $\theta_i~(i\geq 2)$ is the neutrino flux normalization, $\hat{\theta}_i$ and $\hat{\hat{\theta}}_i$ are the component of the MLE $\boldsymbol{\theta}^*$ and $\hat{\boldsymbol{\theta}}$ , respectively, and M is the number of neutrino sources. The binned likelihood has been shown as \forref{eqn:Likelihood}, in which $s_i = s_i(\sigma_0)$ and $b_i = \sum_j b_i^j \theta_j ~(j\geq 2)$ should be the expected WIMP events and neutrino events in the $i^{th}$ bin, respectively. Here, $b_i^j$ represents for the expected neutrino events from the source labelled as $i$. More information on the uncertainties associated with the neutrino flux can be found in Table I of Ref.~\cite{OHare:2020lva}. Then the test statistic can be formulated as follows:
\begin{equation}
\label{eq:q0}
q_{0} =
\left\{ \! \! \begin{array}{ll}
               - 2 \ln \lambda(0)
               & \quad \hat{\theta}_1 \ge 0 \;, \\*[0.3 cm]
               0 & \quad \hat{\theta}_1 < 0  \;.
              \end{array}
       \right. 
\end{equation}
Upon closer examination, it is discernible that the test statistic, denoted by $q_0$, measures the discrepancy between two hypotheses. This discrepancy is shown to increase when the actual observations deviate from the background-only model. According to the definition of discovery limits, the $3~\sigma$ significance of signals corresponds to a p-value $p_0 = 0.0027$. The p-value $p_0$ is defined by $p_0=\int^\infty_{q_{obs}} f(q_0|H_0)dq_{obs}$, where $f(q_0|H_0)$ is the distribution of the statistic under the assumption of $\Hzero$ being true, and is believed to follow a distribution of~\cite{Wilks:1938dza,Chernoff:1954eli,Cowan:2010js}. Thus, the $3~\sigma$ significance corresponds to $q_{obs} = 9$.~\footnote{The significance $Z = \Phi^{-1}(1-p_0)$, where $\Phi^{-1}$ is the quantile of the standard Gaussian~\cite{Cowan:2010js}.}
We also demonstrate it by utilizing Asymptotic-Analytic method, and the numerical analysis reveals the existence of only one non-zero diagonal entry within $\Lambda$ with a value of 1. Therefore, the test statistic $q_0\approx (\sum_i a_i Y_i)^2$, where $\sum_i a_i^2 = 1$ and $Y_i \sim \mathscr{N}(0,1)$, indicating that $q_0$ should follow the chi-square distribution. However, by incorporating the positive condition for the signal strength, the distribution of $q_0$ should be $\frac{1}{2}[ \delta(0)+\chi^2_1]$. The detailed reason of such alteration can be found in Appendix~\ref{appendix2}.

In defining the discovery limits, $P\%$ C.L. is often requested, with previous literature using $P\%=90\%$~\cite{Billard:2011zj,Billard:2013qya} and recent studies opting for $P\%=50\%$~\cite{OHare:2020lva,AristizabalSierra:2021kht}. The $P\%$ C.L. indicates the percentage of experiments in which the discovery is significant, i.e., $\int^\infty_{q_{obs}} f(q_0|H_1)dq_{obs}=P\%$. To obtain the distribution of $q_0$ when $\Hone$ is real, one typically needs to generate numerous pseudodata from MC simulations and perform time-consuming computations to find the MLE. In high-statistics analyses, however, Asimov dataset~\cite{Cowan:2010js} provides a time-saving way to compute the median of $q_0$ when $\Hone$ is real. Thus, a slight modification of the C.L. value, leading to a minor quantitative difference~\cite{OHare:2020lva}. Additionally, Asymptotic-Analytic method reveals that $q_0$ follows a non-central chi-square distribution, denoted as $\chi^2_1(\phi)$, with one DOF and a non-central parameter $\phi$. Note that we can disregard the positive condition for $\theta_1$, because most of $\theta^*_1$ and $\hat{\theta}_1$ are positive in our case where  $q_{obs}=9$ and $\int^\infty_{q_{obs}} f(q_0|H_1)dq_{obs}=P\%$. \zblc{Surprisingly, according to \forref{eqn:VandFSimple} and \forref{eqn:phiSimple}, Asymptotic-Analytic method also afford the analytic form of $\phi$:
\begin{equation}\label{eqn:phi1}
\phi= \left. \sum_i \frac{s_i^2}{v_i} - \sum_{j k}\left[\sum_i \frac{s_i b_i^j}{v_i}\right]_j \left[\sum_i \frac{b_i^j b_i^k}{v_i}+\frac{\delta^j_k}{\sigma_k^2} \right]^{-1}_{~~~~jk} \left[\sum_i \frac{s_i b_i^j}{v_i}\right]_k \right|_{\boldsymbol{\theta}=1}  \,,
\end{equation}
where the quantities enclosed by the square bracket $[\cdots]$ with one and two subscripts represent vectors and matrices, respectively. Besides, $[\cdots]^{-1}$ represents the inverse of a matrix, and $\delta^j_k$ is the Kronecker delta symbol.}

To demonstrate the effectiveness of our new methods, we reproduce the neutrino floor defined by discovery limits and make a comparison with the result from APPEC report~\cite{Billard:2021uyg} based on MC method. Note that the neutrino floor we discuss here relies on the choice on the detect threshold and the exposure. More details can be found in Appendix~\ref{MCCheck}. Consequently, Asymptotic-Analytic and Quasi-Asimov dataset method are effective, except that sometimes they fails on account of insufficient statistics. Moreover, Asymptotic-Analytic technique furnishes a means to assess the influences attributable to individual parameters. As an illustration, the neutrino floor at $m_\chi = 5.5$~GeV, which will be elaborated upon later, is primarily dictated by the $^8$B neutrino flux.

To avoid the setting of somewhat arbitrary experimental configurations and give the neutrino floor a single consistent interpretation in statistics, a new definition~\cite{OHare:2021utq} on the neutrino floor has been proposed and widely recognized. Since the neutrino floor relies on the experimental configurations, it is better appreciated that neutrinos should present a "fog": a region of the parameter space where a clear distinction between signal and background is challenging. Define $n$ as the gradient of a discovery limit cross section $\sigma_{DL}$ with respect to the exposure $N$: $n=-(d \ln \sigma_{DL}/d \ln N)^{-1}$, which is also called the "opacity" of the neutrino fog. The index $n$ alters as $N$ increases. Briefly, when $N$ is small, i.e., the background-free case, $n=1$. Then, as $N$ increases, the case becomes the Poissonian background subtraction: $n=2$. If we have a larger $N$, WIMP signals are covered by fluctuations of neutrino backgrounds, so that $n>2$. That means it is difficult to detect the WIMP signal when we only increase the exposure. Therefore, it is more effective to find another way, like the directional detection strategy, to search for WIMPs. Eventually, as $N$ gets large enough, the intrinsic difference between signals and backgrounds help us discover the signal, and $n=2$ starts to return. In the next section,this evolution of $n$ can be explained analytically with Asymptotic-Analytic method. The neutrino floor is defined as the boundary of the neutrino fog, which  marks the transition from statistical to systematical limits. 
\begin{figure}[!t]
	\centering
	\includegraphics[width=0.75\linewidth]{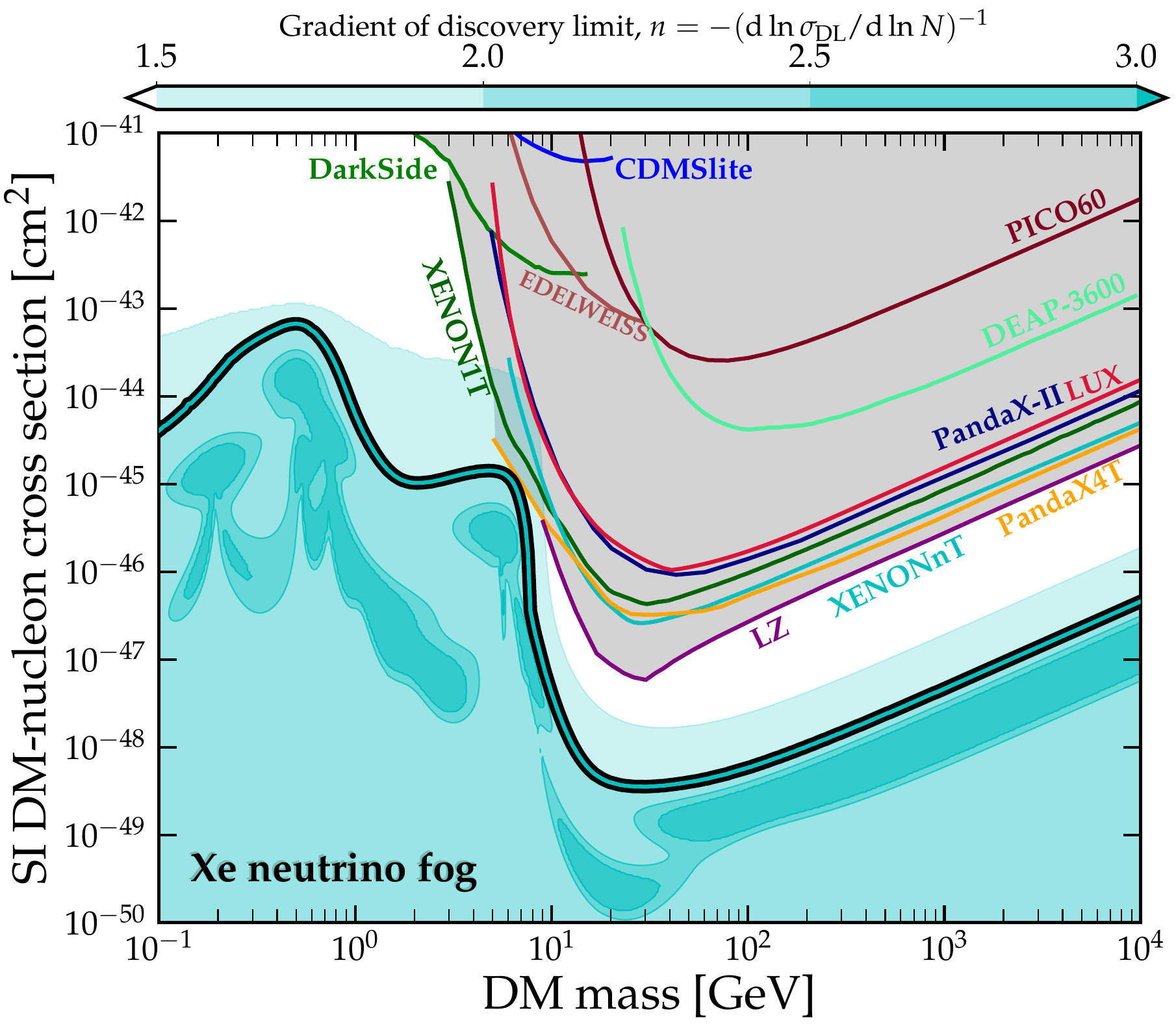}
	\caption{\label{fig:MyNeutrinoFog} The neutrino fog  for the xenon experiment is presented in the cyan contour map, while the solid line represents the neutrino floor for $n=2$. Besides, the latest excluded space is shaded gray~\cite{PandaX-II:2017hlx,XENON:2018voc,PICO:2017tgi,DarkSide:2018kuk,EDELWEISS:2016nzl,DarkSide:2018bpj,LUX:2016ggv,DEAP:2019yzn,SuperCDMS:2018gro,XENON:2023sxq,Liu:2023dig,LZ:2022ufs} (all experimental limits except the latest exlusion limits from XENONnT and PandaX4T are taken from Ref.~\cite{OHare:2021utq}).}
\end{figure}

As illustrated in \figref{fig:MyNeutrinoFog}, we employ Quasi-Asimov dataset method to showcase the neutrino floor and fog. The outcomes from Quasi-Asimov dataset method perfectly reproduce the neutrino floor and fog~\cite{OHare:2021utq,Akerib:2022ort}, albeit with minor discrepancies. In terms of computational efficiency, our methodology outperforms Asimov dataset, taking only about 10 seconds to calculate the neutrino fog on the same computer, as opposed to the half-hour computation required by Asimov dataset. Additionally, Quasi-Asimov dataset method exhibits superior computational stability. More details and figures can be found in our public code~\cite{ourcode}. Note that we follow the calculation techniques outlined in Ref.~\cite{OHare:2021utq}. Thus, events are binned for the recoil spectra in the logarithmic scale between $10^{-4}$ keV and 200 keV. We choose the former detector threshold for the purpose of mapping the neutrino floor down to $m_\chi$ = 0.1~GeV, which actually does not impact the height of the limit at other masses. 

\begin{figure}[!t]
	\centering
	\includegraphics[width=\linewidth]{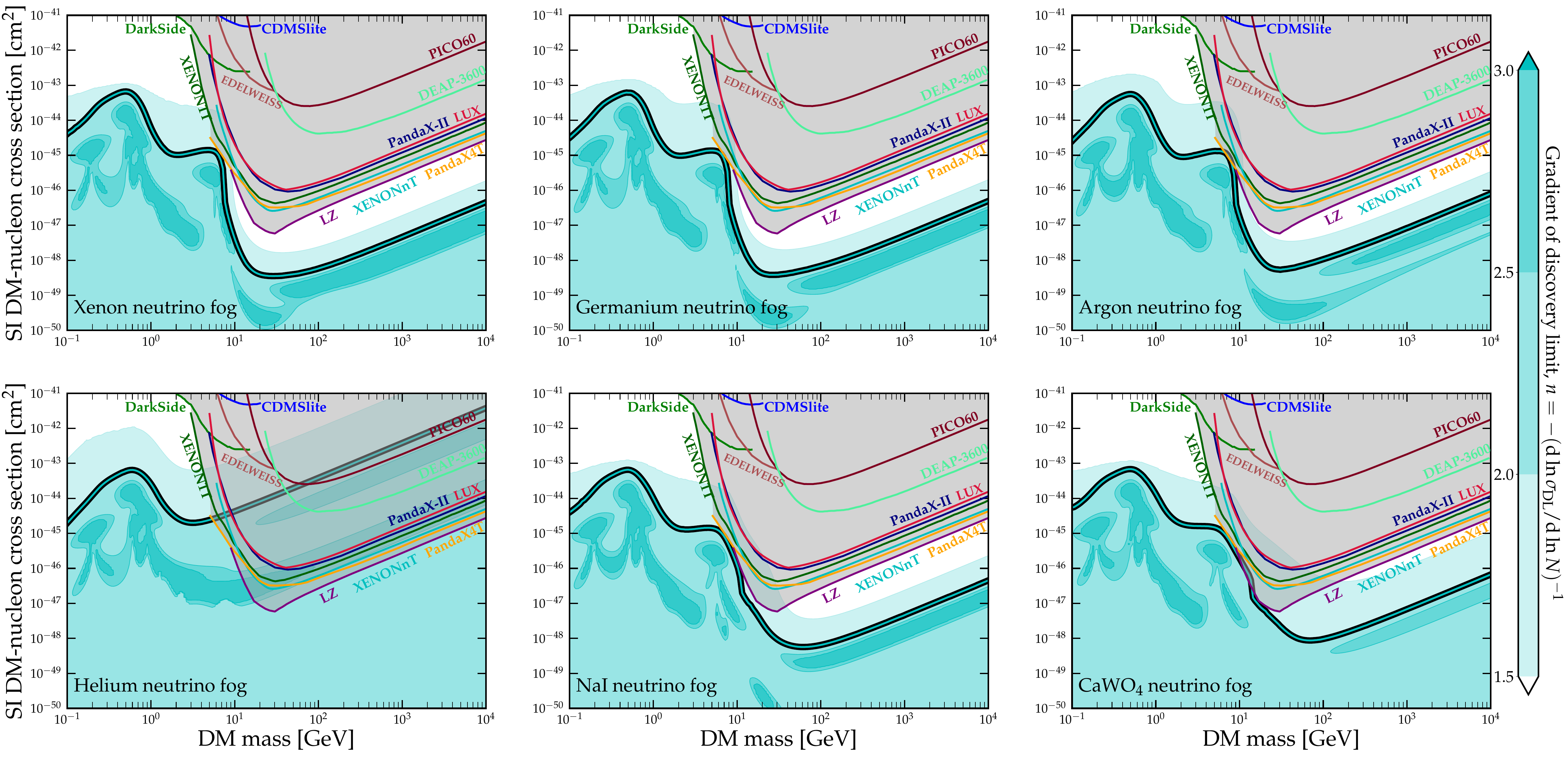}
	\caption{\label{fig:DifferentTargets} The neutrino fogs for six different targets in the direct detection experiments are presented in the cyan contour map, while the solid line represents the neutrino floor for $n=2$. Besides, the latest excluded space is given by the shaded gray region.}
\end{figure}

Moreover, we also present neutrino fogs for six different targets in the DM direct detection experiments. As shown in \figref{fig:DifferentTargets}, neutrino fogs for Xenon, Germanium and Argon targets are similar, while the neutrino fog for the Helium target is quite different and the neutrino floor for the high WIMP mass immerges in the excluded space, since its mass is too low to gain the sensitivity for WIMP with higher mass. Neutrino fogs for composite targets also resemble that derived from Xenon, Germanium and Argon targets, despite of some small differences. For instance, the region around $m_\chi\approx 20$~GeV for the NaI target is more shallow, which indicates that experiments using the NaI target can detect a weaker signal generated by a WIMP of about 20~GeV.

\subsection{Analytical Interpretation on Sensitivity Curves}
\label{Analytic}
When the sample size or exposure is sufficiently large, Asymptotic-Analytic and Quasi-Asimov dataset methods produce consistent results with each other. Therefore, the evolution of the discovery limit cross section $\sigma_\text{DL}$ with respect to $N$ can be analytically explained. For the sake of simplicity, we restrict our attention to the $^8$B neutrino source and $m_\chi=5.5$~GeV. \zblb{As discussed before, the non-central parameter can be obtained from \forref{eqn:SpecificPhi}, 
% \begin{equation} \label{eqn:SpecificPhi}
% \phi= \sum_i \frac{s_i^2}{s_i+b_i} - \frac{(\sum_i \frac{s_i b_i }{s_i+b_i})^2}{\sum_i \frac{b_i^2 }{s_i+ b_i}+\frac{1}{\sigma_\nu^2}} \,,
% \end{equation}
where the expected WIMP events are proportional to $\sigma_\text{DL}$ and $N$: $s_i \propto \sigma_\text{DL} N$, the expected background events also have a linear correlation with $N$: $ b_i \propto N$, and the uncertainty $\sigma_2=\sigma_\nu$, which stands for the uncertainty of the $^8$B neutrino flux.}

\begin{figure}[!t]
	\centering
	\includegraphics[width=0.75\linewidth]{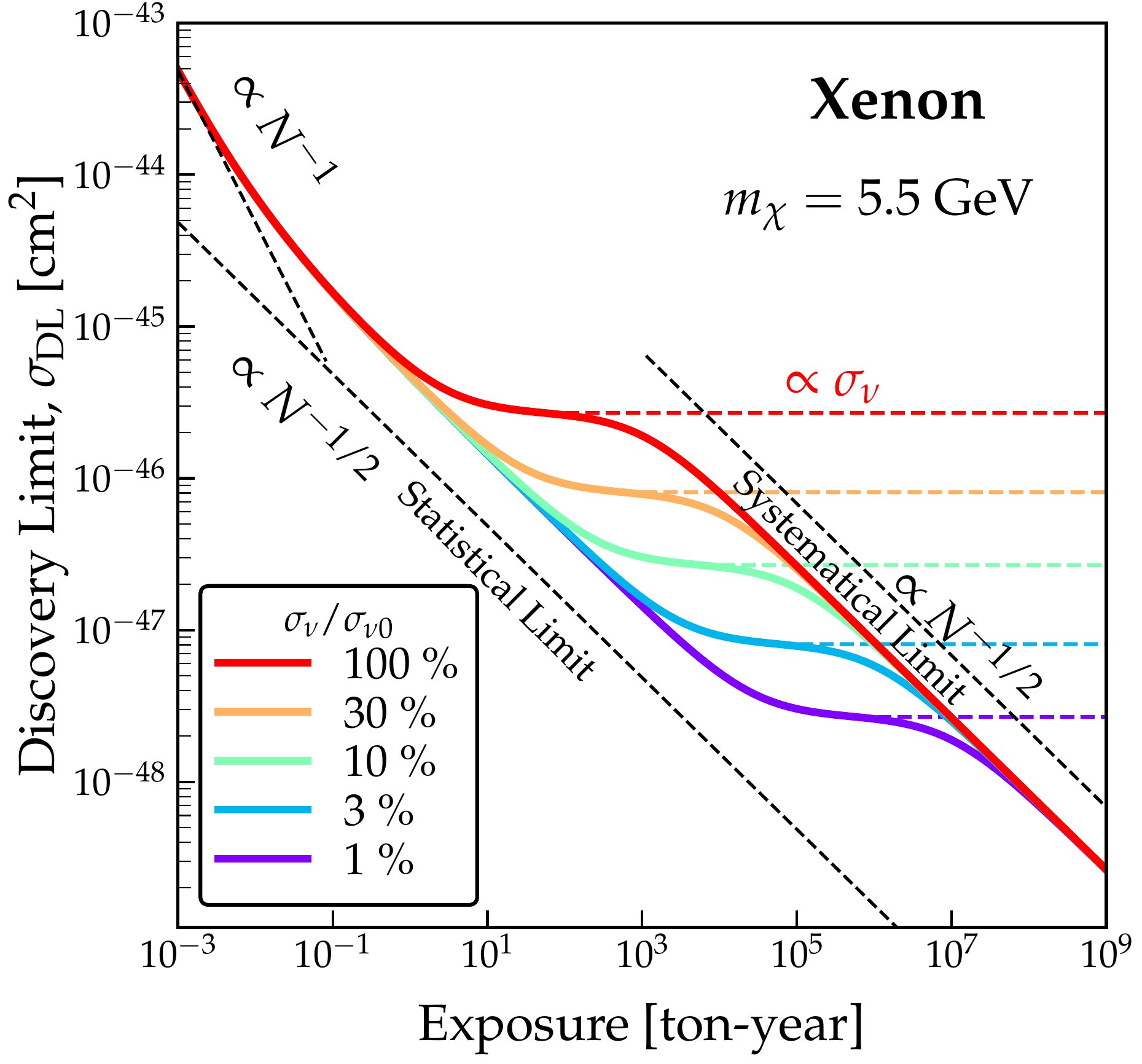}
	\caption{\label{fig:DLCurves} The sensitivity curves about the evolution of the discovery limit cross section with respect to an exposure for $m_\chi=5.5$~GeV. Lines in different colors represent the sensitivity curve with different uncertainties of the $^8$B neutrino flux. \zbld{Here we assume the standard uncertainty of the $^8$B neutrino flux $\sigma_{\nu 0}=2\%$.}}
\end{figure}

Initially, $N<1$, and WIMP signals dominate. Because $s_i\gg b_i$ and $\frac{1}{\sigma_\nu^2}$ dominates in the denominator of the second term in \forref{eqn:SpecificPhi}, the non-central parameter $\phi\approx \sum_i \frac{s_i^2}{s_i+b_i} \approx \mathcal{O}(s) $. Thus, to maintain the invariance of $\phi$, $\sigma_\text{DL} \propto N^{-1}$. As $N$ increases and $\sigma_\text{DL}$ decreases, $s_i < b_i$  while the second term is still controlled by $\frac{1}{\sigma_\nu^2}$. Consequently, we obtain $\phi\approx \sum_i \frac{s_i^2}{s_i+b_i} \approx \mathcal{O}(s^2/b)$, and $\sigma_\text{DL} \propto N^{-\frac{1}{2}}$. Here we come to the Poissonian background subtraction and the statistical limit as illustrated in \figref{fig:DLCurves}.

As $N$ increases further, the signal is lost within the background: $s_i \ll b_i$. Then we have:
\begin{equation}
\phi \approx \sum_i \frac{s_i^2}{b_i} - \frac{(\sum_i s_i)^2}{\sum_i b_i} (1-\frac{1}{\sigma_\nu^2 \sum_i b_i}) \,.
\end{equation}
If there is only one bin for the data or the background mimics the signal: $s_i \propto b_i$, the leading term is the one with $\sigma_\nu$ and $\phi \approx \frac{(\sum_i s_i)^2}{\sigma_\nu^2 (\sum_i b_i)^2} \approx \mathcal{O}(s^2/b^2)$. Hence $\phi$ remains constant as $N$ increases: $\sigma_\text{DL} \propto N^{0}$, which causes the flatness in \figref{fig:DLCurves}. Moreover, since $\phi \approx 1/\sigma_\nu^2$ in this case, a smaller value of the uncertainty $\sigma_\nu$ corresponds to a lower $\sigma_\text{DL}$. Thus, $\sigma_\text{DL} \propto \sigma_\nu$ for the flatness of the curves in  \figref{fig:DLCurves}.
On the other hand, if there are some slight differences between the signal and background even though they are very similar, the term with $\sigma_\nu$ can be neglected and we have $\phi \sim \mathcal{O}(s^2/b)$. Finally, we return to $\sigma_\text{DL} \propto N^{-\frac{1}{2}}$ and reach the systematical limit as shown in \figref{fig:DLCurves}.

\section{Additional Nuisance Parameters}
\label{ANP}
\subsection{\zbld{Weak mixing angle uncertainty}}
\label{WMA}
The weak mixing angle has been well measured at the Z-pole, while it leaves a sizable uncertainties at the low energy (the maximal momentum transfer $q_\text{max}\lesssim 200$~MeV). It might be determined by \cevns process. Conversely, the weak mixing angle also exerts an influence on the response of neutrinos in the DM detector, so we shall explore its impact on the neutrino fog in this section. The weak mixing angle is rather flat over the low energy range, so the neutrino background $b_i\propto Q_{W}^2$, where $Q_{W}=N-(1-4\sin^2\theta_W)Z$ is the weak charge~\cite{Kumar:2013yoa,Erler:2004in,AristizabalSierra:2021kht}. Note that $N$ and $Z$ are the number of neutrons and protons in the target nuclei. As in Ref.~\cite{AristizabalSierra:2021kht}, we take the central values $\sin^2\theta_W=0.2387$ with a 10\% uncertainty. For simplicity, we assign $\theta_{M+1}$ to account for the uncertainty: $Q_{W}=N-(1-4\theta_{M+1}\sin^2\theta_W)Z$, where $M$ represents the number of nuisance parameters from neutrino fluxes.
In this case, $v_i(\boldsymbol{\theta})=\theta_1 s_i + \sum_{2\leq j\leq M} \theta_j b_i^j(\theta_{M+1})$. As the discussion in Section~\ref{AA}, the background-only hypothesis $\Hzero$ corresponds to $\boldsymbol{\theta}^0:~\theta_1^0=0, ~\theta_j^0=1(2\leq j\leq M+1)$, while $\Hone$ corresponds to $\boldsymbol{\theta}^1:\theta_j^1=1(1\leq j\leq M+1)$. 

Observed that $\dot{l}(\boldsymbol{\theta}^\prime)_{M+1}=\frac{8\sin^2\theta_W Z}{Q_w} \sum_j\dot{l}(\boldsymbol{\theta}^\prime)_{j}~(2\leq j \leq M)$ by making use of $\frac{\partial v_i}{\partial \theta_{M+1}}=\frac{8\sin^2\theta_W Z}{Q_w} \sum_j 
 b_i^j$, there are only $M$ independent components in the normal variate $\mathbf{Z}$. Thus, $\mathbf{V}$ has a dimension of $M\times M$, while $\mathscr{F}$ has a larger dimension of $(M+1)\times (M+1)$, as shown in \forref{eqn:complicate}. To solve this problem, we employ a matrix $\mathbf{O}$ to eliminate $\dot{l}(\boldsymbol{\theta}^\prime)_{M+1}$, where 
 $$
\mathbf{O}=\left( 
\begin{array}{c}
     \mathbf{I}_{M\times M}  \\
      0,~\underbrace{\frac{8\sin^2\theta_W Z}{Q_w},\cdots,\frac{8\sin^2\theta_W Z}{Q_w}}_{M}
\end{array}
\right)\,,
$$
where $\mathbf{I}_{M\times M}$ is a diagonal matrix with a dimension of $M\times M$. Since $\boldsymbol{\delta}=0$ is $\Hzero$ is true and $\boldsymbol{\delta}=\{1,~\overbrace{0,\cdots,0}^{M}\}$ if $\Hone$ is true, we can use $\mathbf{O}$ to obtain $Z$ with the correct dimension: $Z \sim \mathscr{N}(\mathbf{V}^{-\frac{1}{2}} \mathbf{O} (\boldsymbol{\mu}+\mathscr{F} \boldsymbol{\delta}), I)$. Therefore, \forref{eqn:StatisticDistribution} is rewritten as:
 \begin{equation}
-2\ln \lambda \approx \mathbf{Z}^T \mathbf{V}^{\frac{1}{2}} \mathbf{O}^T[\mathscr{F}^{-1}-\mathbf{H}]\mathbf{O}
\mathbf{V}^{\frac{1}{2}} \mathbf{Z} \,.
 \end{equation}
According to \forref{eqn:EandV} we have:
\begin{equation}\label{eqn:complicate}
\begin{aligned}
&\mu_\alpha = \sum_{i} \frac{1}{2}\frac{\partial^2 v_i}{\partial \theta_{M+1}^2} \frac{\sigma_2^2}{v_i} \frac{\partial v_i}{\partial \theta_\alpha},~1\leq \alpha \leq M+1\,,\\
&\mathscr{F}_{\alpha \beta} = \left\{\begin{array}{cc}
   -\sum_{i} \frac{1}{2}\frac{\partial^2 v_i}{\partial \theta_{M+1}^2} \frac{\sigma_{M+1}^2}{v_i} \frac{\partial^2 v_i}{\partial \theta_\alpha \partial \theta_\beta} 
+\sum_{i} \frac{1}{v_i}(1+\frac{1}{2}\frac{\partial^2 v_i}{\partial \theta_{M+1}^2} \frac{\sigma_{M+1}^2}{v_i}) \frac{\partial v_i}{\partial \theta_\alpha} \frac{\partial v_i}{\partial \theta_\beta} \,, & \alpha=1~\text{or}~\beta=1\,, \\
    -\sum_{i} \frac{1}{2}\frac{\partial^2 v_i}{\partial \theta_{M+1}^2} \frac{\sigma_{M+1}^2}{v_i} \frac{\partial^2 v_i}{\partial \theta_\alpha \partial \theta_\beta} 
+\sum_{i} \frac{1}{v_i}(1+\frac{1}{2}\frac{\partial^2 v_i}{\partial \theta_{M+1}^2} \frac{\sigma_{M+1}^2}{v_i}) \frac{\partial v_i}{\partial \theta_\alpha} \frac{\partial v_i}{\partial \theta_\beta}
+\delta^\alpha_\beta \frac{1}{\sigma_{\alpha}^2}, & ~2\leq \alpha,\beta \leq M+1\,,
\end{array}\right.
\\
&V_{\alpha \beta} = \sum_{i j k} \left(\frac{1}{v_i} \delta^i_j+\frac{1}{2}\frac{\partial^2 v_i}{\partial \theta_{M+1}^2} \frac{\sigma_{M+1}^2}{v_i^2}\delta^i_j+
 \frac{\partial v_i}{\partial \theta_k} \frac{\partial v_j}{\partial \theta_k} \frac{\sigma_k^2}{v_i v_j}-
\frac{1}{4} \frac{\partial^2 v_i}{\partial \theta_{M+1}^2}
\frac{\partial^2 v_j}{\partial \theta_{M+1}^2} \frac{\sigma_{M+1}^4}{v_i v_j}
 \right) \frac{\partial v_i}{\partial \theta_\alpha} \frac{\partial v_j}{\partial \theta_\beta},~1\leq \alpha,\beta \leq M,~
\end{aligned}
\end{equation}
where the only non-vanishing second derivative $\frac{\partial^2 v_i}{\partial \theta_{M+1}^2}$ is considered, instead of summing over all the nuisance parameters given in \forref{eqn:EandV}. This case is more complicate than the case where only the uncertainties from neutrino fluxes are considered.

Through the numerical computations discussed in Section~\ref{AA}, the test statistic still follows the asymptotic $\frac{1}{2}[ \delta(0)+\chi^2_1]$ distribution if $\Hzero$ is real, while the distribution of $q_0$ is distorted by the variation of $\sin^2\theta_W$ if $\Hone$ is real. When $\Hone$ is real, Asymptotic-Analytic method reveals that there is only one non-zero diagonal element in $\Lambda$ with its value at 1, which means that there is only one $\chi^2$ variate, and the coefficient is not unity. Thus, the distribution of $q_0$ can be expressed as $q_0 \sim a \chi^2_1(\phi)$. In Appendix~\ref{MCCheck}, Monte Carlo realisations and Asymptotic-Analytic method for some benchmark points are shown. It should be noted that although there may be slight deviations between our results and the Monte Carlo realizations, our method remains effective.

\begin{figure}[!t]
	\centering
	\includegraphics[width=0.45\linewidth]{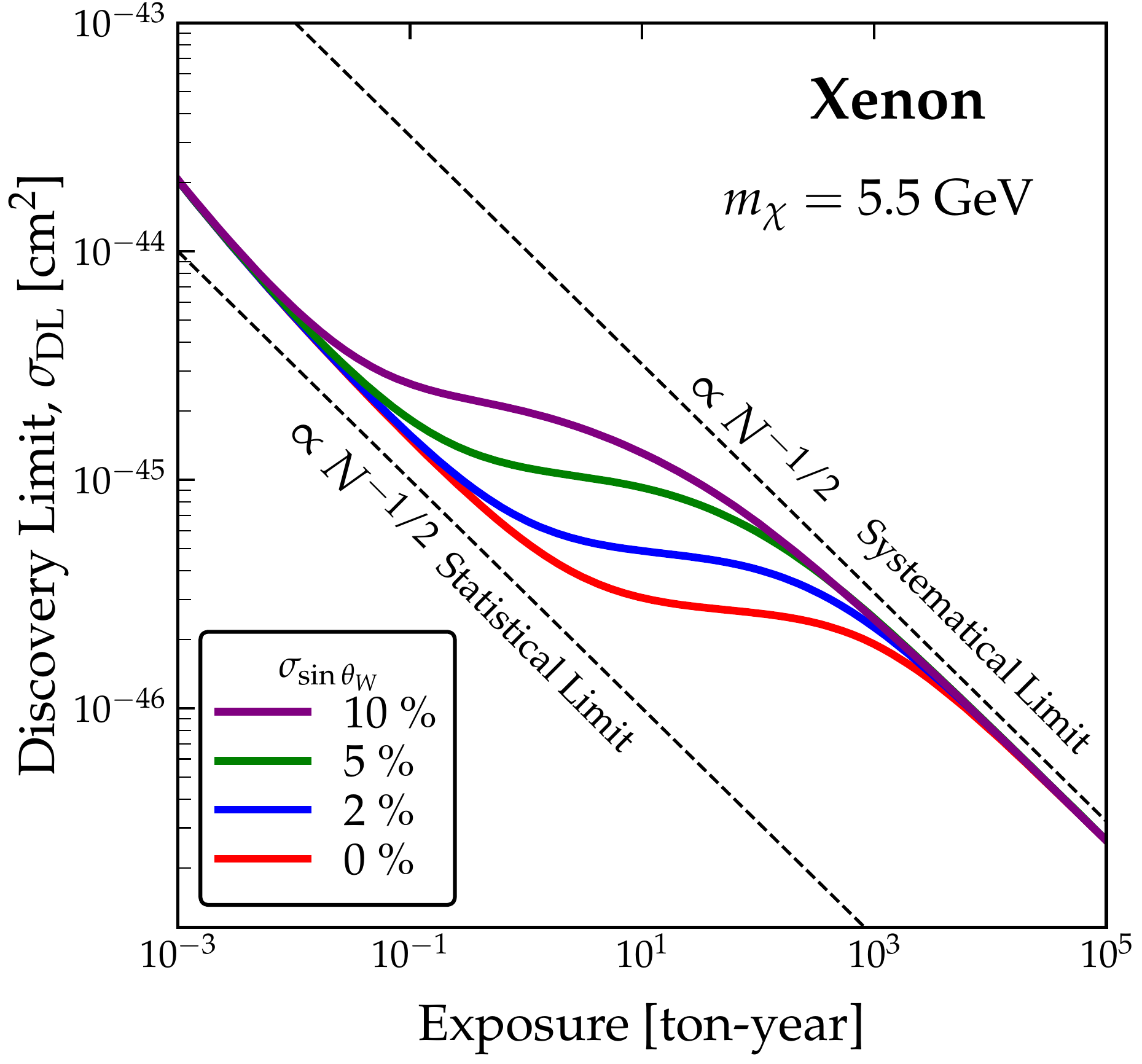}
    \includegraphics[width=0.45\linewidth]{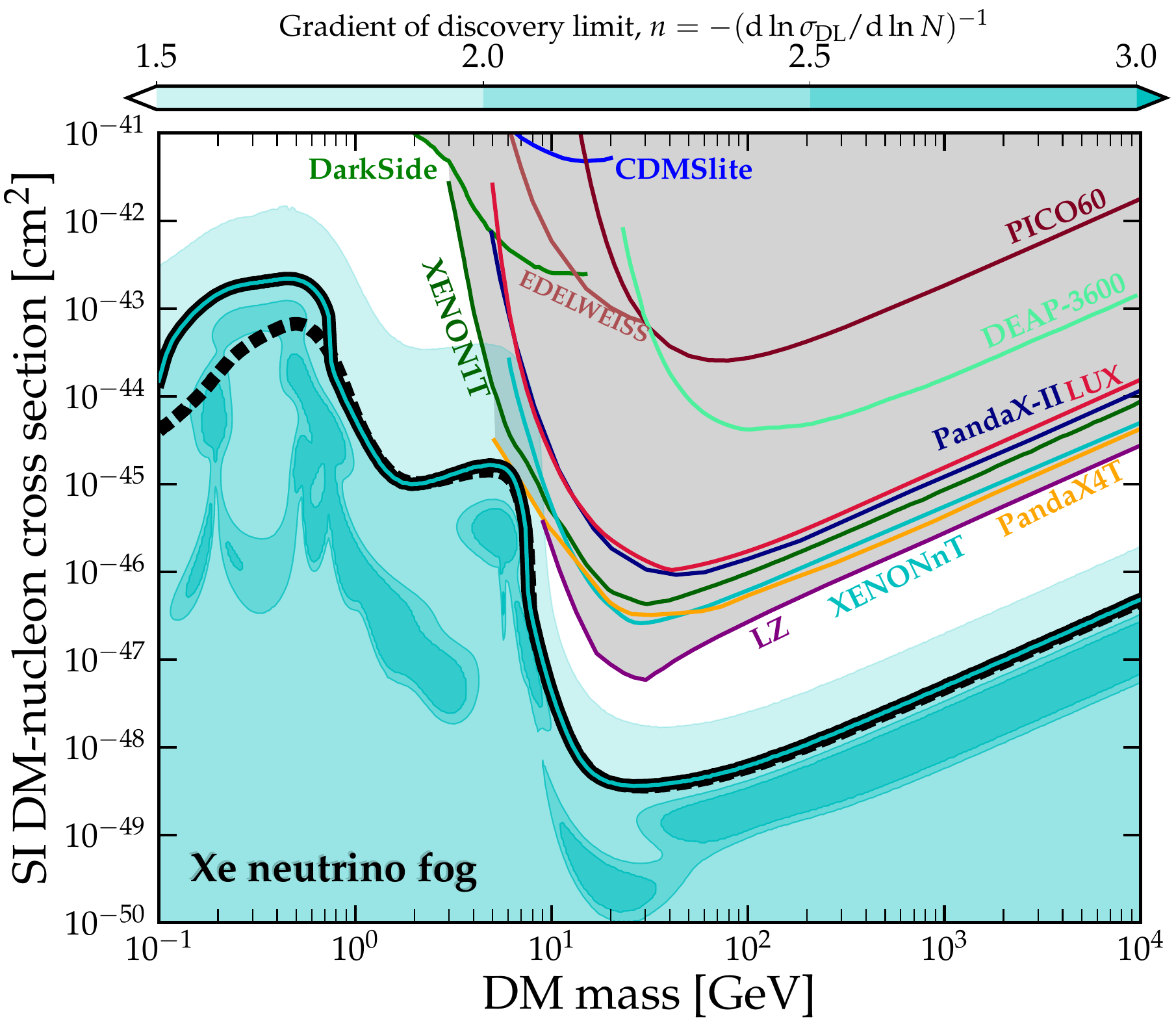}
	\caption{\label{fig:WMA}The sensitivity curves about the evolution of $\sigma_\text{DL}$ with respect to an exposure for $m_\chi=5.5$~GeV, involving the uncertainty of $\sin^2\theta_W$ (left). Lines in different colors represent the sensitivity curve with different uncertainties of $\sin^2\theta_W$. The uncertainty of the $^8$B neutrino flux is set to $2\%$.
 The neutrino fog  for the xenon experiment is presented in the cyan contour map,involving the uncertainty of $\sin^2\theta_W$ (right). The solid line represents the neutrino floor for $n=2$, while the dashed black line represents the neutrino floor without considering the uncertainty of $\sin^2\theta_W$.}
\end{figure}

Similar to the scenario we present in Section~\ref{neutrino}, Quasi-Asimov dataset method is utilized to obtain the discovery limit cross section. As shown in the left panel of \figref{fig:WMA}, the uncertainty of $\sin^2\theta_W$ significantly affect the discovery limit. It is evident that a greater uncertainty leads to a larger $\sigma_\text{DL}$ at the same exposure. However, as the exposure increases, these curves converge to the same systematic limit. Note that we only consider the $^8$B neutrino as depicted in \figref{fig:DLCurves}, for the purpose of comparison. Moving to the right panel of \figref{fig:WMA}, the neutrino fog considering the uncertainty of $\sin^2\theta_W$ is presented. One can see that the uncertainty of $\sin^2\theta_W$ remarkably modifies the neutrino fog shown in \figref{fig:MyNeutrinoFog} over the low mass range ($m_\chi \lesssim 1$~GeV), while the region over the higher mass range remains unchanged. This fact confirms the result in Ref.~\cite{AristizabalSierra:2021kht}. The neutrino floor is elevated when compared to the neutrino floor without considering the uncertainty of $\sin^2\theta_W$, thereby demonstrating the consistency of the new definition of the neutrino floor.

\subsection{Astrophysical uncertainty}
\label{LSR}
As elaborated in Section \ref{statistic}, a quandary regarding Asimov dataset still persists. In the context of the signal discovery,  when some extra nuisance parameters are only implicated in $s_i$, \zbld{the zero signal strength $\theta^0_1=0$ forces the MLEs for $L(\theta^0_1,
\hat{\hat{\theta}}_2,\dots,\hat{\hat{\theta}}_M, \hat{\hat{\theta}}_{M+1}) $ to be indistinguishable from the scenario where the extraneous parameters are absent. Here, an extra nuisance parameter $\theta_{M+1}$ is only involved in the signal, i.e., $s_i=\theta_1 s_i(\theta_{M+1})$.} This outcome is contrary to our expectations, and fortunately, our Asymptotic-Analytic method is adept at resolving this predicament.

For the sake of convenience, we shall confine our attention to the velocity of the local standard of rest (LSR) $v_0$ and its uncertainty as the extra nuisance parameter.  According to the previous investigation~\cite{OHare:2016pjy}, we simply surmise that $v_0$ is subject to a normal distribution, with a mean of 220 km/s and a standard deviation of 50 km/s. Analogous to our handling of neutrino fluxes, the velocity is scaled to unity, and its standard deviation is 21.2\%. \zbld{Denoted by $\theta_{M+1}$ as in Section~\ref{WMA}, this nuisance parameter is distinguished from $\theta_j$ $(2\leq j \leq M)$, which pertains to the neutrino flux. In this case, $v_i(\boldsymbol{\theta})=\theta_1 s_i(\theta_{M+1}) + \sum_{2\leq j\leq M} \theta_j b_i^j$ no longer suffices the simple linear form as $\theta_{M+1}$ modifies the shape of WIMP spectrum. Consequently, we must re-examine \forref{eqn:StatisticDistribution} and \forref{eqn:EandV} to elicit the asymptotic distribution of the test statistic. In this case, $\mathbf{V}$ and $\mathscr{F}$ has the same dimension of $(M+1)\times (M+1)$, which can be directly computed by \forref{eqn:complicate} where the dimension of $\mathbf{V}$ should be changed into $(M+1)\times (M+1)$. As the discussion in Section~\ref{AA}, the background-only hypothesis $\Hzero$ corresponds to $\boldsymbol{\theta}^0:~\theta_1^0=0, ~\theta_j^0=1(2\leq j\leq M+1)$, while $\Hone$ corresponds to $\boldsymbol{\theta}^1:\theta_j^1=1(1\leq j\leq M+1)$.} Comparatively, utilization of Asimov dataset reveals that the discovery limits, with or without consideration of $v_0$, coincide, a result which is at odds with the findings in the existing literature~\cite{OHare:2016pjy}. 

Instead of utilizing the laborious Monte Carlo realisations, our Asymptotic-Analytic method is available to address this case. Note that we adhere to the 90\% C.L. for the discovery limit to illustrate the feature of our method. In this instance, as $\hat{\hat{\theta}}_{M+1}$ has been firmly established to 1, we are in fact in pursuit of the MLE for $L(0,
\hat{\hat{\theta}}_2,\dots,\hat{\hat{\theta}}_M, 1) $ which corresponds to the case of the two parameters of interest. When $\Hzero$ is real,  the computational outcome manifests that the test statistic still follows the asymptotic $\frac{1}{2}[ \delta(0)+\chi^2_1]$ distribution. While for $\Hone$ is real, the test statistic is asymptotic distributed to the $a_1 \chi^2_1(\phi_1)+a_2 \chi^2_1(\phi_2)$ distribution, \zbld{where $a_i$ and $\phi_i$ ($i=1,2$) can be obtained from the numerical solution as presented in Section~\ref{AA}. Numerically, the asymptotic distribution of $q_0$ can be obtained by performing
the inverse Fourier transformation on its characteristic function.} To show our method's effectiveness, we compare the test statistic's distribution from Monte Carlo realisations and Asymptotic-Analytic method for some benchmark points, which can be found in Appendix~\ref{MCCheck}.

\begin{figure}[!t]
	\centering
	\includegraphics[width=0.45\linewidth]{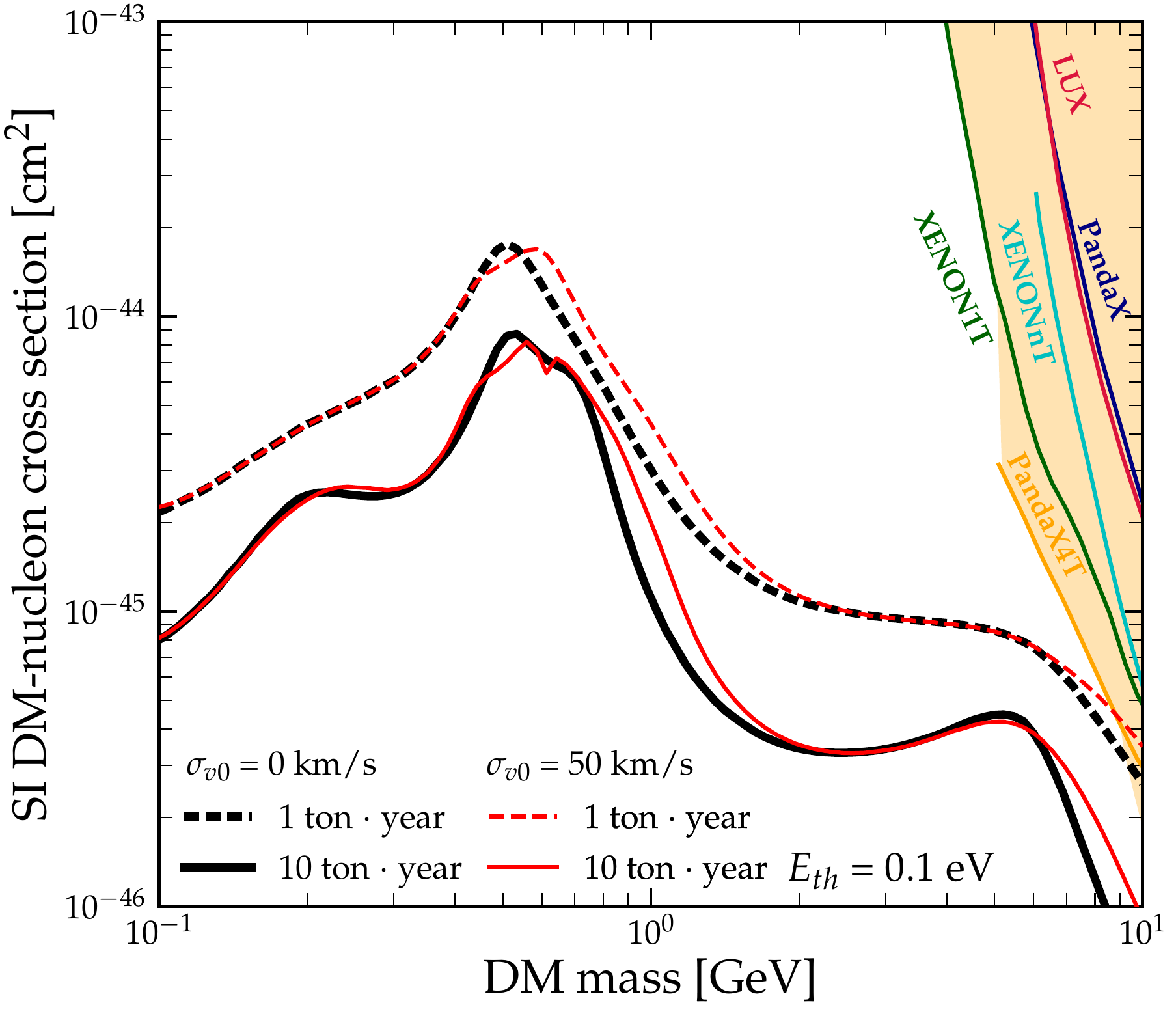}
    \includegraphics[width=0.45\linewidth]{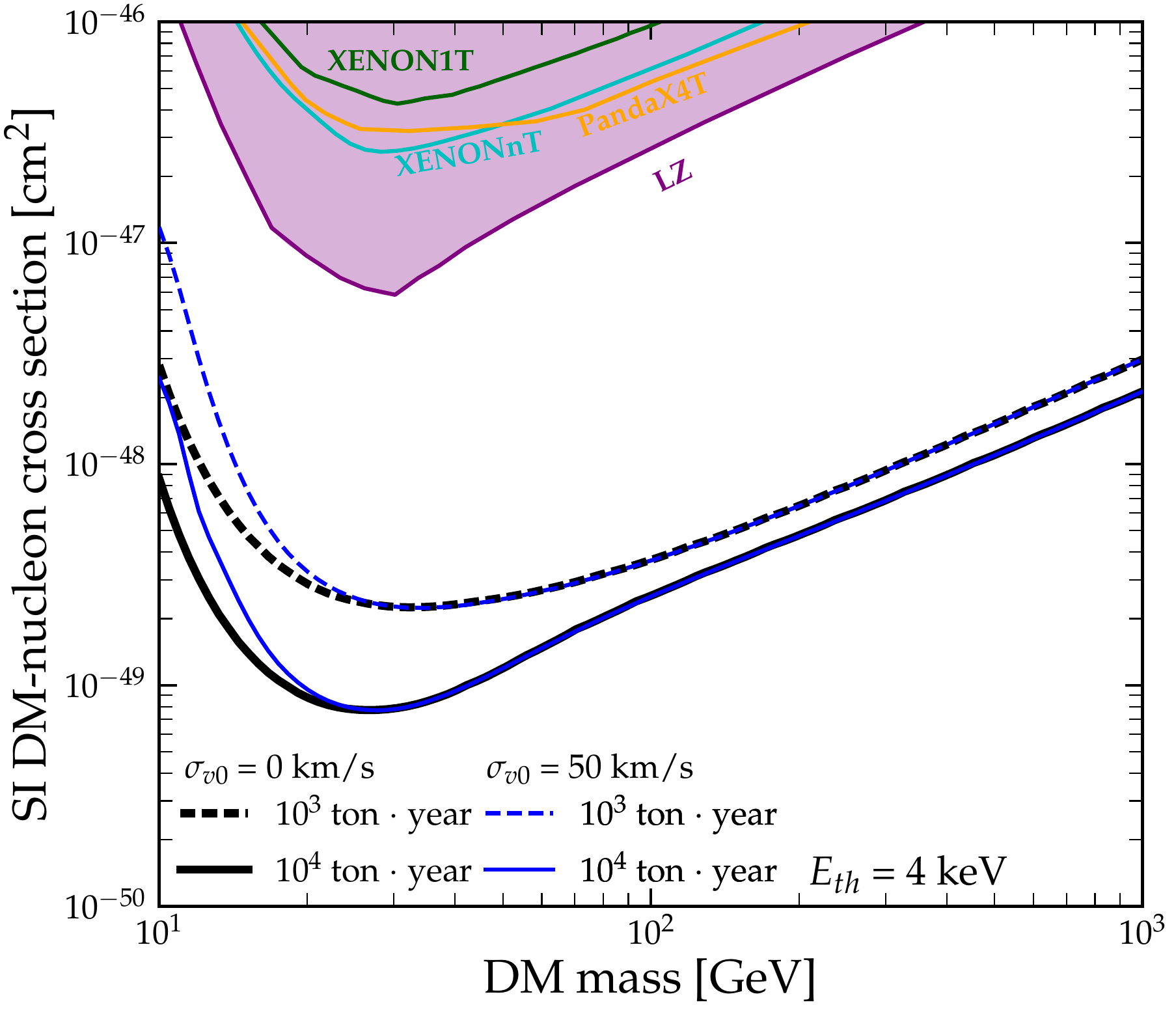}
	\caption{\label{fig:LSR}Discovery limits with and without considering the extra parameter from the velocity of LSR, which are displayed in colored and black lines. The left panel corresponds to the case where the threshold equals to 0.1~eV with two different exposures: 1~\tonyear and 10~\tonyear. The right panel corresponds to the case where the threshold is 4~keV with two different exposures: $10^3$~\tonyear and $10^4$~\tonyear. The solid and dashed lines stand for lower and higher exposures, respectively. The latest excluded limits are also shown.}
\end{figure}

\zbld{Nevertheless, it is worth noting that for extremely large exposures, the numerical solution becomes unreliable, since the matrices $\mathbf{V}$ and $\mathscr{F}$ are no longer positive definite. Consequently, it is not possible to accurately present the neutrino fog considering the uncertainty of $\theta_2$, while discovery limits for two benchmark scenarios are presented instead. Note that the covariance matrix $\mathbf{V}$ should be always positive definite, and a positive definite $\mathscr{F}$ guarantees the stationary point is the minimum.}
As illustrated in \figref{fig:LSR}, we implement Asymptotic-Analytic method to calculate the discovery limit that pertains to the nuisance parameter from $v_0$ on the WIMP parameter space. Additionally, we have provided the outcome that involves only the nuisance parameters from neutrino fluxes for the purpose of comparison. 
In the low mass region, $m_\chi \lesssim 10$~GeV, we adopt the optimal threshold of 0.1 eV as previously utilized, since it facilitates the mapping of the limit down to $m_\chi = 0.1$~GeV. Besides, we choose the exposure of 1~\tonyear and 10~\tonyear. Generally speaking, introducing an extra nuisance parameter should raise the limit, as the region around $m_\chi\approx 1$~GeV in \figref{fig:LSR}. However, certain interesting things have been observed around the 0.5 GeV and 6 GeV, where the limits involving an extra parameter become lower on the contrary. We demonstrate the feature with MC realizations which can be found in Appendix~\ref{MCCheck}. The similar phenomenon has been discussed in Ref.~\cite{OHare:2016pjy}, while it still needs further investigation. For other regions, no significant changes on the limit are observed, while the distribution of the test statistic has been distorted. For further information, please refer to Appendix~\ref{MCCheck}.

For the larger mass range $m_\chi \gtrsim 10$~GeV, we adopt a more realistic threshold at 4 keV, while it needs larger exposures to gain enough statistics. So we choose the exposure of $10^3$~\tonyear and $10^4$~\tonyear so that events from atmosphere and DSNB are sufficient for  CEvNS in the same nuclear recoil energy region stand out as clear signals. It can be seen from \figref{fig:LSR} that the introduction of $v_0$ significantly raises up the limit around $m_\chi\approx 10$~GeV, while the limit for $m_\chi \gtrsim 30$~GeV. This fact indicates that the effect of the astrophysical uncertainties can not be ignored especially when we analyze the neutrino floor for the WIMP mass around 10~GeV.

\section{Conclusions and Outlook}
\label{conclusion}
In this paper, we have investigated the asymptotic behaviour of the profile binned likelihood ratio test statistic, in which the likelihood is constructed from different variables and pull terms, drawing inspiration from the seminal works of Wilk and Ward~\cite{Wilks:1938dza,Wald1943TestsOS}. Based on our findings, we have proposed two new methods: Asymptotic-Analytic method, which can provide analytical results for large statistics, handle situations involving some specific nuisance parameters and affords a way to determine the most relevant parameters in the statistical analysis; and Quasi-Asimov dataset method, which is similar to but faster than Asimov dataset. We make a comparison on the computational speed, accuracy of results and extensibility for four methods in Tab.~\ref{tab:methods}. Our proposed methods are not only applicable for the neutrino floor and fog, but also feasible for other studies utilizing Asimov dataset in experimental analysis and phenomenology. Moreover, the current methodology in statistics will pave the way to scrutinize the origin of tiny discrepancy in a comparison of theoretical predictions and experimental data from the DM experiments where new physics might be hidden~\cite{zbl:2023}.
\begin{table}[!t]
\label{tab:methods}
\caption{Comparison on the computational speed, accuracy of results and extensiblility for four methods.}
\begin{center}
\begin{tabular}{c|c|c|c|c}
\hline
                           & Speed            & Accuracy                        & Extensible         & References                                                     \\ \hline \hline
Monte Carlo simulation     & $\sim$1 day      & $\checkmark$                    & $\checkmark$       & \cite{Billard:2013qya, Ruppin:2014bra}                       \\ \hline
Asimov dataset             & $\sim$30 minutes & $\checkmark$                    & sometimes $\times$ & \cite{AristizabalSierra:2021kht,OHare:2020lva,OHare:2021utq} \\ \hline
Asymptotic-Analytic Method & $\sim$10 seconds & $\checkmark$ for big statistics & $\checkmark$       & This work                                                          \\ \hline
Quasi-Asimov dataset       & $\sim$10 seconds & $\checkmark$                    & sometimes $\times$ & This work                                                           \\ \hline
\end{tabular}
\end{center}
\end{table}

We have employed our newly proposed methods on the neutrino floor and fog. By utilizing Quasi-Asimov dataset, we have achieved near-perfect reproduction of the neutrino floor and fog~\cite{OHare:2021utq,Akerib:2022ort} with a computational speed that is two orders of magnitude faster and improved stability. \zblb{Likewise, we also consider the uncertainty of the weak mixing angle in the context of the neutrino fog, and verify the consistency of the new definition of the neutrino floor.} On the other hand, Asymptotic-Analytic method provides an analytical formula to quantitatively explain the evolution of the discovery limit cross section with exposure, and it offers a solution for cases involving astrophysical uncertainties that cannot be dealt with by Asimov dataset. Additionally, our methods are capable and effective when considering more degrees of freedom in the context of the neutrino floor, such as the detector efficiency, the resolution and new physics beyond the standard model. Furthermore, MC pseudo-experiments can be boosted by only considering the most relevant parameters, which can be obtained by Asymptotic-Analytic method, from neutrino fluxes.

Our method with the decent derivation with an approximation to speed up the computation has worked very well in the context of the neutrino floor and fog, as demonstrated by the aforementioned numerical calculations. Nevertheless, there are a few drawbacks to our proposed methods that we should give warnings and address carefully. 
Firstly, Asymptotic-Analytic method may be out of service when the sample size is too small. However, Quasi-Asimov dataset method remains effective in obtaining the median of the test statistic, as Asimov dataset does. Secondly, Asymptotic-Analytic method requires small uncertainties; otherwise, higher-order corrections are necessary, which can complicate matters.

\section*{Acknowledgement}
We appreciate Dr. Jia-Jie Ling for useful discussions. This project was supported in part by National Natural Science Foundation of China under Grant No. 12075326 and Fundamental Research Funds for the Central Universities (23xkjc017), Sun Yat-sen University. 

\appendix

\section{The asymptotic formula of the test statistic}
\label{appendix1}
Assuming the log likelihood function $l(\boldsymbol{\theta})$ can be approximated as the quadratic function around $\hat{\boldsymbol{\theta}}$ or $\boldsymbol{\theta}^\prime$ in the limit of large samples, $l(\boldsymbol{\theta}^*)$ is expanded about $\hat{\boldsymbol{\theta}}$:
\begin{equation*}
l(\boldsymbol{\theta}^*) \approx l(\hat{\boldsymbol{\theta}}) + \dot{l}(\hat{\boldsymbol{\theta}}) (\boldsymbol{\theta}^*-\hat{\boldsymbol{\theta}}) + \frac{1}{2} (\boldsymbol{\theta}_n^*-\hat{\boldsymbol{\theta}})^T \ddot{l}(\hat{\boldsymbol{\theta}})  (\boldsymbol{\theta}^*-\hat{\boldsymbol{\theta}}) +\mathscr{O}(\frac{1}{\sqrt{N}})\,,
\end{equation*}
\zbl{where $E(\ddot{l}(\hat{\boldsymbol{\theta}}))$ is the expectation of the second derivative of the log likelihood function at $\hat{\boldsymbol{\theta}}$, and $N$ represents the sample size and the last term can be safely neglected when $N$ is large enough. As a rule of thumb, the total event number $N$ should be $\mathscr{O}(100)$. Otherwise, the approximation here might be out of service.} Assuming that $\hat{\boldsymbol{\theta}}$ is close enough to $\boldsymbol{\theta}^\prime$ for the large sample case, we have $\ddot{l}(\hat{\boldsymbol{\theta}}) \approx \ddot{l}(\boldsymbol{\theta}^\prime)$. According to the Lyapunov CLT, $\ddot{l}(\boldsymbol{\theta}^\prime)$ comprising numerous variate can asymptotically approximate to its expectation $E(\ddot{l}(\boldsymbol{\theta}^\prime))$, i.e., $\ddot{l}(\boldsymbol{\theta}^\prime)\approx E(\ddot{l}(\boldsymbol{\theta}^\prime))$. Thus, with $\dot{l}(\hat{\boldsymbol{\theta}})=0$ from the MLE condition and defining $\mathscr{F}=-E(\ddot{l}(\boldsymbol{\theta}^\prime))$, we have:
\begin{equation*}
-2\ln \lambda = 2(l(\boldsymbol{\theta}^*)-l(\hat{\boldsymbol{\theta}})) \approx (\boldsymbol{\theta}^*-\hat{\boldsymbol{\theta}})^T \mathscr{F}  (\boldsymbol{\theta}^*-\hat{\boldsymbol{\theta}}) \,.
\end{equation*}

\zbl{Thanks to the Lyapunov CLT, we can infer that the first derivative of the log likelihood function at $\boldsymbol{\theta}^\prime$, i.e., $\dot{l}(\boldsymbol{\theta}^\prime)$ follows a multivariate normal distribution, denoted by $\mathscr{N}(\boldsymbol{\mu}, \mathbf{V})$. Here, $\boldsymbol{\mu}=E(\dot{l}(\boldsymbol{\theta}^\prime))$ represents the expectation vector, and $ \mathbf{V}=var(\dot{l}(\boldsymbol{\theta}^\prime))$ is the variance matrix.} Since $\dot{l}(\boldsymbol{\theta}^\prime)\sim \mathscr{N}(\boldsymbol{\mu}, \mathbf{V})$, we better relate $(\boldsymbol{\theta}^*-\hat{\boldsymbol{\theta}})$ with $\dot{l}(\boldsymbol{\theta}^\prime)$ to obtain the asymptotic distribution. Expand $\dot{l}(\boldsymbol{\theta}^*)$ about $\hat{\boldsymbol{\theta}}$:
\begin{equation}
\label{eqn:proof1}
\dot{l}(\boldsymbol{\theta}^*) \approx  \dot{l}(\hat{\boldsymbol{\theta}}) +E(\ddot{l}(\boldsymbol{\theta}^\prime)) (\boldsymbol{\theta}^*-\hat{\boldsymbol{\theta}}) \approx -\mathscr{F} (\boldsymbol{\theta}^*-\hat{\boldsymbol{\theta}}) \,.
\end{equation}
Besides, denote the deviation from $\boldsymbol{\theta}^\prime$ to $\boldsymbol{\theta}^0$ as $\boldsymbol{\delta}= \boldsymbol{\theta}^\prime-\boldsymbol{\theta}^0$, and expand $\dot{l}(\boldsymbol{\theta}^*)$ about $\boldsymbol{\theta}^\prime$:
\begin{equation}
\label{eqn:proof2}
\dot{l}(\boldsymbol{\theta}^*) \approx 
\dot{l}(\boldsymbol{\theta}^\prime) -\mathscr{F}(\boldsymbol{\theta}^*-\boldsymbol{\theta}^0)  +\mathscr{F} \boldsymbol{\delta}\,.
\end{equation}
Let
$$
\mathscr{F}\equiv\left[\begin{array}{cc}
\mathbf{G}_1 (r \times r) & \mathbf{G}_2 (r \times(k-r))\\
\mathbf{G}_2^T((k-r) \times r) & \mathbf{G}_3((k-r) \times(k-r))
\end{array}\right], \quad
\mathbf{H}\equiv\left[\begin{array}{cc}
0 & \mathbf{0} \\
\mathbf{0} & \mathbf{G}_3^{-1}
\end{array}\right] \,.
$$
\zblb{where $\mathbf{G}_1,~\mathbf{G}_2,~\mathbf{G}_3$ are the block matrices inside $\mathscr{F}$, and $r,~k$ represent their dimensions.}
Because the last k-r components of $\dot{l}(\boldsymbol{\theta}^*)$  and the first r components of  $(\boldsymbol{\theta}^*-\boldsymbol{\theta}^0)$ are zero, multiplying $\mathbf{H}$ on the left side of \forref{eqn:proof2} we have $\mathbf{H} \dot{l}(\boldsymbol{\theta}^*) =0$ and then:
\begin{equation}
0\approx \mathbf{H}\dot{l}(\boldsymbol{\theta}^\prime) - (\boldsymbol{\theta}^*-\boldsymbol{\theta}^0) + \mathbf{H}\mathscr{F}\boldsymbol{\delta}
\,.
\end{equation}
Substitute it into \forref{eqn:proof2} and return to \forref{eqn:proof1}, we have:
\begin{equation}
\label{eqn:proof3}
\boldsymbol{\theta}^*-\hat{\boldsymbol{\theta}} \approx -\mathscr{F}^{-1}
\dot{l}(\boldsymbol{\theta}^*) \approx -(\mathscr{F}^{-1}- \mathbf{H}) (\dot{l}(\boldsymbol{\theta}^\prime)+\mathscr{F} \boldsymbol{\delta}) \,.
\end{equation}
Thus, with $\mathbf{H}\mathscr{F}^{-1}\mathbf{H}=\mathbf{H}$, we finally obtain:
\begin{equation}\label{eqn:proof4}
-2\ln \lambda \approx (\dot{l}(\boldsymbol{\theta}^\prime)+\mathscr{F} \boldsymbol{\delta})^{T}[\mathscr{F}^{-1}-\mathbf{H}](\dot{l}(\boldsymbol{\theta}^\prime)+\mathscr{F} \boldsymbol{\delta})\,.
\end{equation}
\zbl{For convenience, let $Z = \mathbf{V}^{-\frac{1}{2}} (\dot{l}(\boldsymbol{\theta}^\prime)+\mathscr{F} \boldsymbol{\delta}) \sim \mathscr{N}(\mathbf{V}^{-\frac{1}{2}} (\boldsymbol{\mu}+\mathscr{F} \boldsymbol{\delta}), I)$, we obtain \forref{eqn:StatisticDistribution}:
\begin{equation*}
-2\ln \lambda \approx Z^T \mathbf{V}^{\frac{1}{2}}[\mathscr{F}^{-1}-\mathbf{H}]
\mathbf{V}^{\frac{1}{2}} Z \,.
\end{equation*}
}

\section{Evaluating quantities in Asymptotic-Analytic Method with an example}
\label{appendix2}
From \forref{eqn:Likelihood}, \zblc{we obtain the log likelihood and its derivations:
\begin{equation*}
\begin{aligned}
&l(\boldsymbol{\theta}) =
-\sum_{i=1}^N [v_i - n_{i}\ln v_i] - \sum_{j=1}^M \frac{(\theta_j - \theta_j^\prime)^2}{2\sigma_{j}^2} + const.\,,\\
&\dot{l}(\boldsymbol{\theta})_\alpha = \left\{\begin{array}{cc}
    \sum_{i=1}^N [\frac{n_{i}}{v_i}-1] \frac{\partial v_i}{\partial \theta_\alpha}\,, & \alpha=1\,, \\
    \sum_{i=1}^N [\frac{n_{i}}{v_i}-1] \frac{\partial v_i}{\partial \theta_\alpha}- \frac{\theta_\alpha - \theta_\alpha^\prime}{\sigma_\alpha^2}, & 2\leq \alpha \leq M\,,
\end{array}\right.
\\
&\ddot{l}(\boldsymbol{\theta})_{\alpha \beta} = \left\{\begin{array}{cc}
    \sum_{i=1}^N [\frac{n_{i}}{v_i}-1] \frac{\partial^2 v_i}{\partial \theta_\alpha \partial \theta_\beta}  -\sum_{i=1}^N \frac{n_{i}}{v_i^2} \frac{\partial v_i}{\partial \theta_\alpha} \frac{\partial v_i}{\partial \theta_\beta}\,, & \alpha=1~\text{or}~\beta=1\,, \\
    \sum_{i=1}^N [\frac{n_{i}}{v_i}-1] \frac{\partial^2 v_i}{\partial \theta_\alpha \partial \theta_\beta}  -\sum_{i=1}^N \frac{n_{i}}{v_i^2} \frac{\partial v_i}{\partial \theta_\alpha} \frac{\partial v_i}{\partial \theta_\beta} - \delta^\alpha_\beta \frac{1}{\sigma_{\alpha}^2}, & 2\leq \alpha,\beta \leq M\,.
\end{array}\right.
\end{aligned}
\end{equation*}
}

Prior to calculating the values of $\boldsymbol{\mu}$, $\mathbf{V}$ and $ \mathscr{F}$, one can easily compute the  $E(n_i)$ as follows:
\begin{equation*}
E(n_i) = \prod_\alpha\int_\alpha d\theta_\alpha \mathscr{N}(\theta_\alpha,\sigma_\alpha) \sum_j \mathscr{P}(n_i|v_i(\boldsymbol{\theta})) = 
\prod_\alpha\int_\alpha d\theta_\alpha \mathscr{N}(\theta_\alpha,\sigma_\alpha) v_i(\boldsymbol{\theta})
= v_i(\boldsymbol{\theta}) + \sum_\alpha \frac{1}{2}\frac{\partial^2 v_i}{\partial \theta_\alpha^2} \sigma_\alpha^2\,,
\end{equation*}
where only the term with $\mathcal{O}(\sigma_j^2)$ is taken into consideration, assuming that $\sigma_j$ is sufficiently small. However, one can opt for more perturbation orders to achieve a more precise outcome. Analogously, we possess:$E(n_i n_j) = v_i v_j + \sum_\alpha \frac{1}{2}\frac{\partial^2 v_i v_j + \delta^i_j v_j}{\partial \theta_\alpha^2} \sigma_\alpha^2 + \delta^i_j v_j$. 

\zblc{Then we acquire the analytical results of $\boldsymbol{\mu}$, $\mathbf{V}$ and $\mathscr{F}$ in \forref{eqn:EandV}. Especially, in our neutrino floor case $v_i(\boldsymbol{\theta})=\theta_1 s_i + \sum_{j\geq 2} \theta_j b_i^j$, and we have:
\begin{equation}\label{eqn:VandFSimple}
\begin{aligned}
&\mu_\alpha = 0,\quad 1\leq \alpha \leq M\,,\\
&\mathscr{F}_{\alpha \beta} = \left\{\begin{array}{cc}
  \sum_{i} \frac{1}{v_i} \frac{\partial v_i}{\partial \theta_\alpha} \frac{\partial v_i}{\partial \theta_\beta} \,, & \alpha=1~\text{or}~\beta=1\,, \\
   \sum_{i} \frac{1}{v_i} \frac{\partial v_i}{\partial \theta_\alpha} \frac{\partial v_i}{\partial \theta_\beta}
+\delta^\alpha_\beta \frac{1}{\sigma_{\alpha}^2}, & 2\leq \alpha,\beta \leq M\,,
\end{array}\right.
\\
&V_{\alpha \beta} = \sum_{i} \frac{1}{v_i} \frac{\partial v_i}{\partial \theta_\alpha} \frac{\partial v_i}{\partial \theta_\beta} + \sum_{i j k} 
 \frac{\partial v_i}{\partial \theta_k} \frac{\partial v_j}{\partial \theta_k} \frac{\sigma_k^2}{v_i v_j}
\frac{\partial v_i}{\partial \theta_\alpha} \frac{\partial v_j}{\partial \theta_\beta} ,~1\leq \alpha,\beta \leq M\,.
\end{aligned}
\end{equation}
When $\Hzero$ is real: $\boldsymbol{\theta}^\prime = \boldsymbol{\theta}^0$, the numerical solution tell us that there is only one non-zero diagonal element in $\Lambda$ with its value at 1. Therefore,  the statistic $q_0=-2\ln \lambda \approx a_1 Y_1^2, a_1 = 1, Y_1 \sim \mathscr{N}(0,1)$. 
Furthermore, the signal strength $\theta_1$ should be positive in this situation, which leads to the fact that $q_0$ asymptotically follows the  $\frac{1}{2}[\delta(0)+\chi^2_1]$ distribution instead of $\chi^2_1$.} It can be explained by the \forref{eqn:proof3} for $\boldsymbol{\delta}=0$: 
\begin{equation}
\boldsymbol{\theta}^*-\hat{\boldsymbol{\theta}}  \approx -(\mathscr{F}^{-1}- \mathbf{H}) \dot{l}(\boldsymbol{\theta}^\prime)\,,
\end{equation}
where every component of $\boldsymbol{\theta}^*-\hat{\boldsymbol{\theta}}$ is a normal variate. With the positive condition for $\theta_1$, all negative values of $\theta^*_1-\hat{\theta}_1$ are forced to be zero. Consequently, the distribution of the statistic is altered to $\frac{1}{2}[ \delta(0)+\chi^2_1]$.

\zblc{When $\Hone$ is real: $\boldsymbol{\theta}^\prime = \boldsymbol{\theta}^1$, similarly, the numerical solution reveals that there is only one non-zero diagonal element in $\Lambda$ with value of 1. Thus, the statistic $q_0 \approx a_1 Y_1^2, a_1 = 1, Y_1 \sim \mathscr{N}(\sqrt{\phi},1)$, where $\phi$ is:
\begin{equation}
\phi = \boldsymbol{\delta}^T \mathscr{F}[\mathscr{F}^{-1}-\mathbf{H}] \mathscr{F} \boldsymbol{\delta}\,.
\end{equation}
It can be directly obtained from \forref{eqn:proof4}, where $\dot{l}(\boldsymbol{\theta}^\prime)$ is neglected since its mean value $\boldsymbol{\mu}$ is zero in this case. Making use of the only non-vanishing component in $\boldsymbol{\delta}$ is $\delta_1=1$, we perform a more useful form of $\phi$:
\begin{equation}\label{eqn:phiSimple}
\phi = \delta_1 (\mathbf{G}_1-\mathbf{G}_2 \mathbf{G}_3^{-1} \mathbf{G}_2^T ) \delta_1\,,
\end{equation}
where the dimension of $\mathbf{G}_1$ is $(1\times1)$.}

\section{Compared with results from MC method}
\label{MCCheck}
\begin{figure}[!t]
	\centering
	\includegraphics[width=0.75\linewidth]{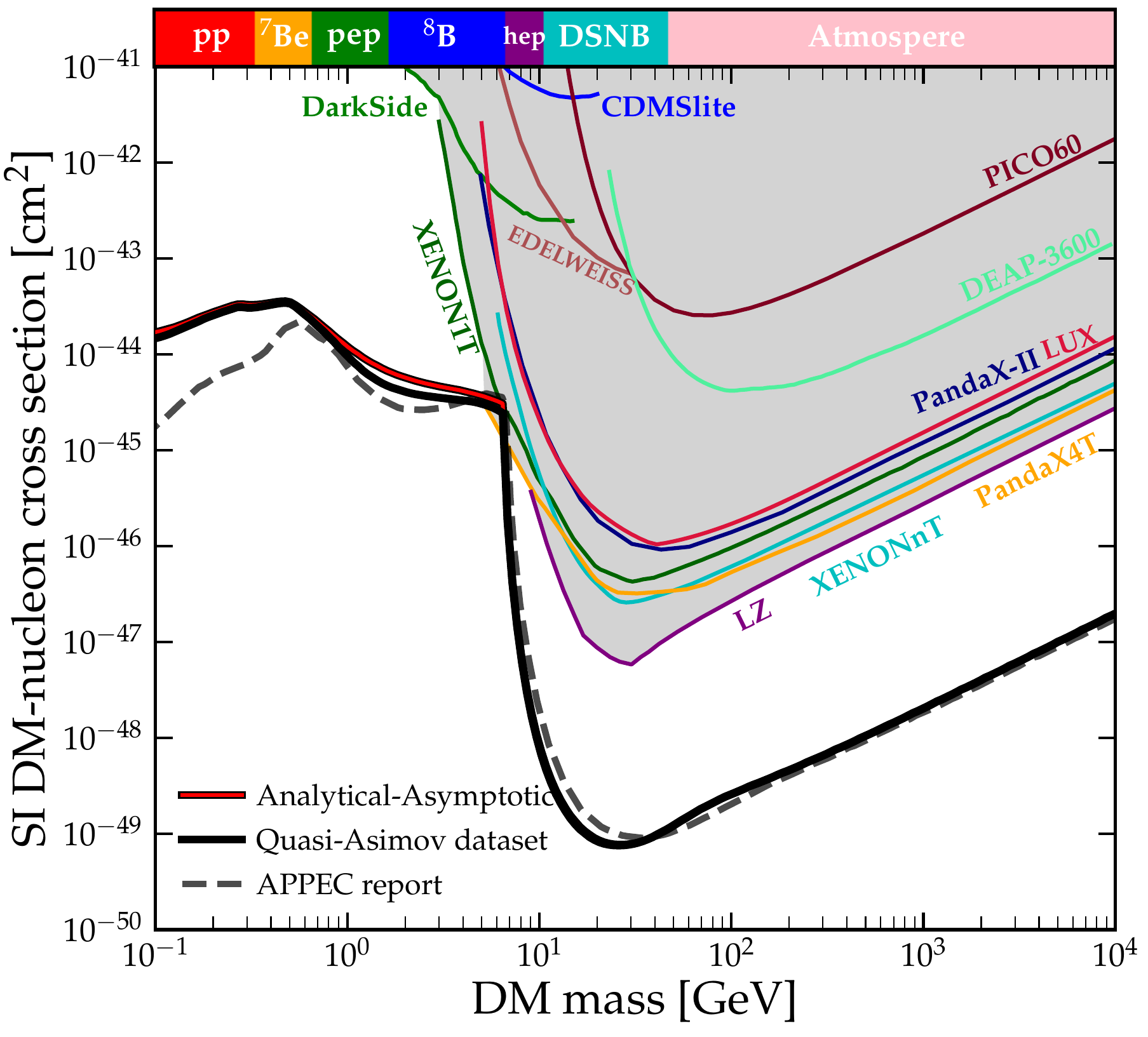}
	\caption{\label{fig:RegeneratingVFloor} Neutrino floors defined by discovery limits in the DM parameter space. The red and black solid lines stand for the results from Asymptotic-Analytic and Quasi-Asimov dataset method, respectively. The black dashed line is taken from APPEC report. The color-coded neutrino sources, causing primary effects on the neutrino floor over the WIMP mass range, is shown above the figure.}
\end{figure}
\begin{figure}[!t]
	\centering
     \includegraphics[width=0.47\linewidth]{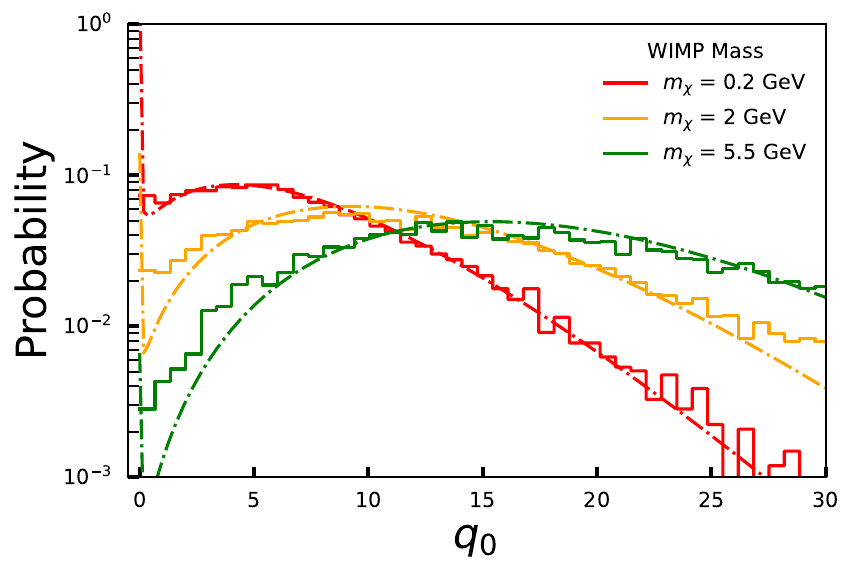}
	\includegraphics[width=0.47\linewidth]{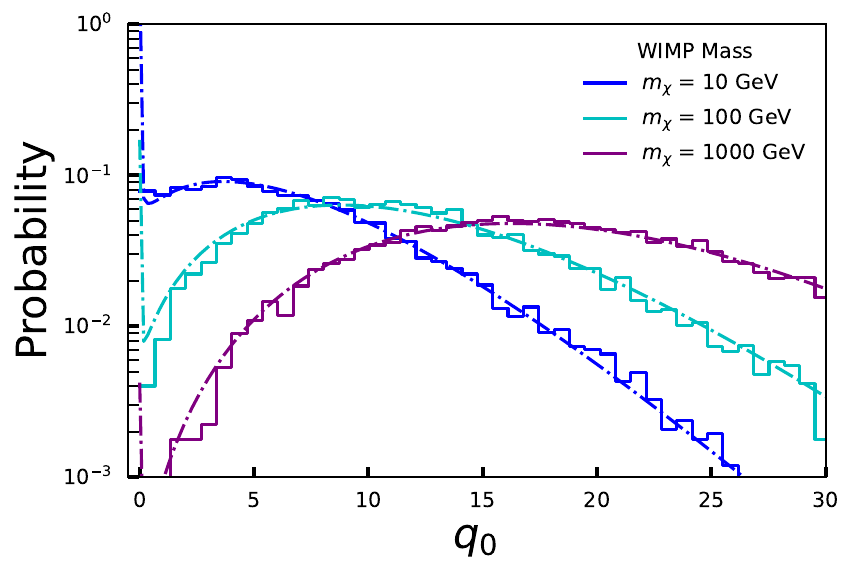}
	\caption{\label{fig:MCNu} Comparison between the test statistic's distribution from the theoretical asymptotic prediction and the 10000 MC pseudo-experiments. The non-central parameter $\phi$ is gained by Quasi-Asimov dataset method. More details can be found in our public codes~\cite{ourcode}.}
\end{figure}
As demonstrated in \figref{fig:RegeneratingVFloor}, two new methods in this study are used to reproduce the neutrino floor, and the results closely match the neutrino floor from APPEC report for $m_\chi \gtrsim 6$~GeV. However, there are some discrepancies between our results and APPEC results, which could be caused by the differences in the threshold settings, neutrino fluxes, and uncertainties. To check this, we select some benchmark points along the neutrino floor and obtain the distribution of the test statistic using MC pseudo-experiments, as shown in \figref{fig:MCNu}. Furthermore, as depicted in \figref{fig:RegeneratingVFloor}, the neutrino floor from Asymptotic-Analytic method is slightly higher than that from Quasi-Asimov dataset method. This is because the sample size is not large enough to invalidate the approximation utilized in our deduction in Section~\ref{statistic}.
\begin{figure}[!t]
	\centering
 \includegraphics[width=0.49\linewidth]{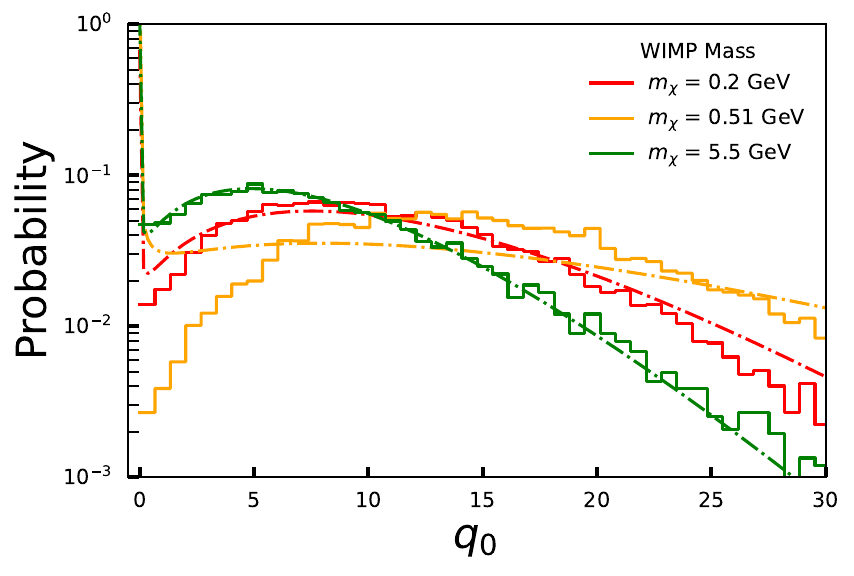}
	\includegraphics[width=0.49\linewidth]{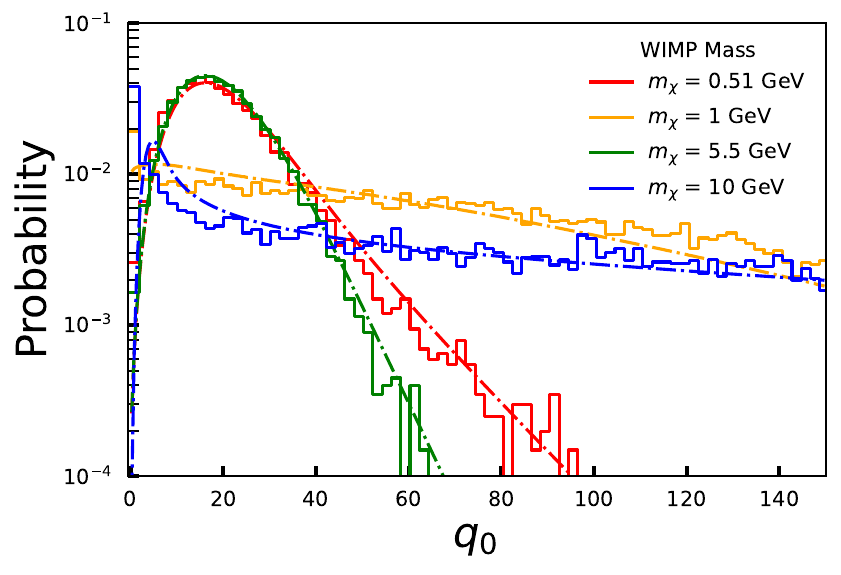}
	\caption{\label{fig:MCLSR} Comparison between the test statistic's distribution from the theoretical asymptotic prediction and 10000 MC pseudo-experiments, where the test statistic involves the weak mixing angle (left) and the nuisance parameter from the velocity of LSR (right).}
\end{figure}

In order to validate our methods' effectiveness, we reproduce the neutrino floor from APPEC report~\cite{Billard:2021uyg} based on MC method. Since the neutrino floor defined by discovery limits depends on the detector configuration, we need to combine several neutrino spectra for different setup to reach the final neutrino floor. The detector threshold can be chosen to be realistic or ideal, while its multiplication with the exposure is assumed to contribute to about 500 neutrino events~\cite{Billard:2013qya}. As shown in \figref{fig:RegeneratingVFloor}, two new methods are utilized to reproduced the neutrino floor and the results closely match the neutrino floor from APPEC report for $m_\chi \gtrsim 6~$ GeV. However, there are some tiny discrepancies between our results and APPEC's result, which might be caused by the differences in the settings on the detector thresholds, neutrino fluxes and their uncertainties. For $m_\chi \lesssim 6$~GeV, since the exposure corresponds to the threshold we choose is not large enough (only 0.018~\tonyear), the distribution of the test statistic somehow deviates from the non-central chi-square distribution. To check this, we have chosen some benchmark points along the neutrino floor and obtained the distribution of the test statistic using MC pseudo-experiments, as shown in \figref{fig:MCNu}. Besides, it can be seen from \figref{fig:RegeneratingVFloor} that the neutrino floor from Asymptotic-Analytic method is slightly higher than that from Quasi-Asimov dataset method. That is because the sample size is not large enough so that the approximation utilized in our deduction in Section~\ref{statistic} becomes invalid.

\zblb{In order to validate the effectiveness of our method as discussed in Section~\ref{WMA}, MC pseudo-experiments are presented here. Note that we fix the exposure at 10~\tonyear, and choose the appropriate cross section to demonstrate our results. As shown in the left panel of \figref{fig:MCLSR}, we take three benchmark points of interest and run 10000 MC psudo- experiments for each point. It can be seen from the left panel of  \figref{fig:MCLSR} that there are some sizable discrepancies between the result of $m_\chi=0.51$~GeV and the corresponding prediction. For $m_\chi=0.1$~GeV and 5.5~GeV, our predictions are in agreement of MC pseudo-experiments.}

We also use MC realizations to confirm our statement in Section~\ref{LSR}. Note that we use the technique described in Section~\ref{AA} to save computational expense by only taking more relevant parameters as inputs. As shown in \figref{fig:MCLSR}, we take four benchmark points of interest and run 10000 MC psudo- experiments for each point. More details can be found in our public codes~\cite{ourcode}. One can see that the results from Asymptotic-Analytic method are consistent with those from MC realizations, despite of tiny discrepancies. For $m_\chi =10$~GeV, we might need higher-order corrections since the standard deviation of 21.2\% is relatively large in this case.

%\newpage
\bibliographystyle{unsrt}
\bibliography{reference}

\end{document}